\documentclass[twocolumn,twocolappendix]{aastex631}
\usepackage{newtxtext,newtxmath,amssymb}
\usepackage[normalem]{ulem}

\renewcommand{\vec}[1]{\pmb{#1}}

\usepackage{calligra}

\begin{document}


\title{Polarization Angle Orthogonal Jumps in Fast Radio Bursts}

\author[0000-0003-4721-4869]{Yuanhong Qu}\thanks{E-mail: yuanhong.qu@unlv.edu}
\affiliation{Nevada Center for Astrophysics, University of Nevada, Las Vegas, NV 89154}
\affiliation{Department of Physics and Astronomy, University of Nevada Las Vegas, Las Vegas, NV 89154, USA}

\author[0000-0002-9725-2524]{Bing Zhang}\thanks{E-mail: bing.zhang@unlv.edu}
\affiliation{Nevada Center for Astrophysics, University of Nevada, Las Vegas, NV 89154}
\affiliation{Department of Physics and Astronomy, University of Nevada Las Vegas, Las Vegas, NV 89154, USA}

\author{Pawan Kumar}\thanks{E-mail: pk@astro.as.utexas.edu}
\affiliation{Department of Astronomy, University of Texas at Austin, Austin, TX 78712, USA}

\begin{abstract}
Recently, polarization angle (PA) orthogonal jumps over millisecond timescales were discovered from three bursts of a repeating fast radio burst source FRB 20201124A by the FAST telescope.  
In general, PA jumps can arise from the coherent or incoherent superposition of two electromagnetic waves, with total polarization fraction remains constant in the former and not in the latter. 
The observations seem to be more consistent with incoherent superposition. The amplitudes of the two orthogonal modes are required to be comparable when jumps occur. 
We provide general constraints on FRB emission and propagation mechanisms based on the data. Physically, it is difficult to produce PA jumps through switching the dominance of the two orthogonal modes within millisecond timescales, 
and a geometric effect due to the source rotation is more plausible.
This requires that the emission region be within the magnetosphere of a spinning central engine, likely a magnetar. 
The two orthogonal modes in different directions can arise when the source rotation brings two independent emission regions with different dominant modes successively into the line-of-sight,
either due to intrinsic radiation mechanisms or the O-mode undergoing a delayed transparency because of the Alfv\'en-O-mode conversion. 
Splitting of emission directions for the two modes due to plasma birefringence is not easy to achieve when the plasma is moving relativistically. For intrinsic radiation mechanisms, curvature radiation always predicts $|E_{\rm X}/E_{\rm O}|\gtrsim1$, and is difficult to produce jumps; whereas inverse Compton scattering can achieve the conversion amplitude ratio $|E_{\rm X}/E_{\rm O}|=1$ to allow jumps to occur under special geometric configurations.
\end{abstract}


\section{Introduction}

Fast radio bursts (FRBs) are bright millisecond-duration astronomical transients predominantly originating from cosmological distances \citep{Lorimer2007,Thornton2013}.
The connection between FRBs and magnetar was confirmed with the detection of FRB 20200428 from one Galactic magnetar SGR 1935+2154 \citep{Bochenek2020,CHIME/FRB2020} implying that at least some FRBs can be produced by magnetars. However, the source origins of FRBs are still subject to debate. 
Additionally, the high brightness temperatures $\sim10^{35}$ K require the radiation mechanisms must be coherent, which are also puzzling. 

Growing observational results can provide essential clues on the radiation mechanisms and properties of emission regions \citep{Zhang2023RMP}. 
One of the dominant feature of repeating FRBs is polarization information.
Interestingly, current data from both repeating and non-repeating FRBs show interesting but puzzling features regarding diverse features of polarization angle (PA).
Observationally, the PA as a function of time within individual bursts either shows a flat (non-evolution) or a varying behavior. Interesting observational facts can be summarized as follows:
\begin{itemize}
\item Flat PA temporal evolution has been detected in most active repeating FRBs (e.g. 
FRB 20121102A 
\citep{Gajjar2018,Michilli2018}, FRB 20180916B \citep{CHIME2019c,Chawla2020,Nimmo2021}, FRB 20190711A \citep{Day2020,PKumar&Shannon2021}, FRB 20190303A and FRB 20190417A \citep{Feng2022}, FRB 20190604A \citep{Fonseca2020}, FRB 20201124A \citep{Xu2021,Jiang22}).
In some of these sources, both flat and varying PA evolution have been observed. In most cases for both flat and varying PAs, high linear polarization with $\Pi_L\gtrsim95\%$ is usually detected.

\item The phenomena of varying PA include two general cases: 
PA swings (regular S-shaped or irregular swing) and PA jumps.
Various regular and irregular PA swings have been observed in some individual bursts of FRB 20180301A \citep{Luo2020nature,Feng2022} and FRB 20220912A \citep{ZhangYK2023}.

For the regular S-shaped PA case, some bursts can be well described by the rotating vector model (RVM) \citep{Radhakrishnan&Rankin1990} commonly invoked to study the evolution of pulsar PA. One example is a non-repeating FRB 20221022A detected by the Canadian Hydrogen Intensity Mapping Experiment (CHIME) \citep{Mckinven2024}, suggesting a natural connection between FRBs and magnetospheric emission similar to pulsar radio emission. 
The observed PA swing can be explained by a misaligned magnetar, 
and small PA swing events are attributed to a nearly aligned magnetar \citep{Beniamini&Kumar2025}.
However, for repeating sources such as FRB 20180301A \citep{Luo2020nature}, no unified RVM model works for all bursts, suggesting a dynamically varying magnetosphere in FRB sources.

\item 
Recently, the first polarization angle ninety-degree jump from FRB 20201124A was detected with Five-hundred-meter Aperture Spherical radio Telescope (FAST) \citep{NiuJR2024,Jiang2024}. 
The PA jumps were detected in three bursts with a transition timescale of several milliseconds. The jumps occurred when the linear polarization degree became minimum and the total polarization degree was not constant.
Both temporal evolution and polarization properties can provide insightful information on intrinsic radiation mechanisms of FRBs.
PA jumps have been widely observed \citep{Manchester1975,Backer1976,Stinebring1984} in radio pulsars,
and different orthogonal modes transition processes have been investigated within the framework of radio pulsars \citep{McKinnon2024}. The detection of such jumps suggests that FRB emission may share a similar radiation or propagation mechanism. Superposition of two orthogonal modes has been proposed to explain the PA jumps in both pulsar radio emission and FRBs \citep{Stinebring1984,NiuJR2024}.

\end{itemize} 

In general, the temporal jump of PA within milliseconds can be attributed to either a physical change of the plasma condition within a millisecond timescale 
or a geometric effect as the fixed line of sight (LOS) sweeps different emission regions due to the rotation of the central engine.
For the latter possibility, emissions from two orthogonal modes should be directed in different directions. This can be achieved via either intrinsic mechanisms or geometric effects. In this paper, we will systematically go through these effects and identify the most plausible scenarios to produce PA jumps in FRBs.

This paper is organized as follows. 
In Section~\ref{sec:wave superposition}, we investigate the general physics of superposition of two waves both coherently and incoherently. 
In Section~\ref{sec:general constraint}, 
we discuss the general constraints from the observations and argue that the PA jumps should be induced by a geometrical effect, as the rotation of the FRB central engine brings different emission regions within its magnetosphere into the observer’s line-of-sight.
In Section~\ref{sec:physical scenarios}, we investigate three scenarios (intrinsic emission mechanisms, A-O-mode conversion, and plasma birefringence) for PA jumps inside the magnetosphere.
Conclusions are summarized in Section~\ref{sec:conclusion} with some discussion.

\section{Physics of two waves superposition}\label{sec:wave superposition}

Observationally, the frequency-time (or waterfall) plots of PA jump events clearly show the dominance of two different modes before and after the PA jumps \citep{NiuJR2024}. This suggests that there are intrinsically two emission modes whose PAs differ by 90$^{\rm o}$, which are defined as orthogonal modes.

In this section, we discuss the general physics of the superposition of two electromagnetic waves with orthogonal PAs (e.g. magnetospheric X-mode and O-mode) through coherent and incoherent superposition. 
The physics of polarization transfer and superposition have been studied in the context of radio pulsars 
\citep[e.g.][]{Cocke&Holm1972,Dyks2017,Dyks2019,Dyks2021,Oswald2023,McKinnon2024}. 
We include some basics here for pedagogical purposes to facilitate follow-up discussions.

Consider two monochromatic waves whose electric fields are defined as
\begin{equation}\label{eq:first wave}
E_{1x}=\varepsilon_1 \cos\theta_1 e^{i\phi_{1x}}, \ E_{1y}=\varepsilon_1 \sin\theta_1 e^{i\phi_{1y}},
\end{equation}
and
\begin{equation}\label{eq:second wave}
E_{2x}=\varepsilon_2 \cos\theta_2 e^{i\phi_{2x}}, \ E_{2y}=\varepsilon_2 \sin\theta_2 e^{i\phi_{2y}},
\end{equation}
where $\varepsilon_{1}$ and $\varepsilon_2$ denote the amplitudes of the two waves, $\theta_1$ and $\theta_2$ denote the angle between the global electric field of each wave and $x$-axis, and $\phi_{ij}$ is the phase for each electric field component with $i=1,2$ and $j=x,y$.
For each wave, the electric field vector is projected in two orthogonal axes that are perpendicular to each other. 
The superposition of the two waves can be calculated based on the principles for either coherent or incoherent superpositions. 

In order to calculate the observed polarization properties, one should calculate the four Stokes parameters defined as \citep{Rybicki&Lightman1979}
\begin{equation}
\begin{aligned}
&I=\frac{1}{2}(E_1^*E_1+E_2^*E_2), \ &&Q=\frac{1}{2}(E_1^*E_1-E_2^*E_2),\\
&U={\rm Re}(E_1^*E_2), \ &&V={\rm Im}(E_1^*E_2).
\end{aligned}
\end{equation}
where $E_1$ and $E_2$ denote two orthogonal electric fields on the plane which is perpendicular to the LOS.
The linear, circular, and the total degree of polarization can be calculated as
\begin{equation}
\Pi_L=\frac{\sqrt{Q^2+U^2}}{I}, \ \Pi_V=\frac{V}{I}, \ \Pi_p=\sqrt{\Pi_L^2+\Pi_V^2}.
\end{equation}
The PA is defined as
\begin{equation}
{\rm PA}=\frac{1}{2}\arctan\frac{U}{Q}.
\end{equation}

In the following, we discuss the two ways of superposition in detail.

\subsection{Coherent superposition}

\begin{figure*}
\begin{center}
\begin{tabular}{ll}
\resizebox{80mm}{!}{\includegraphics[]{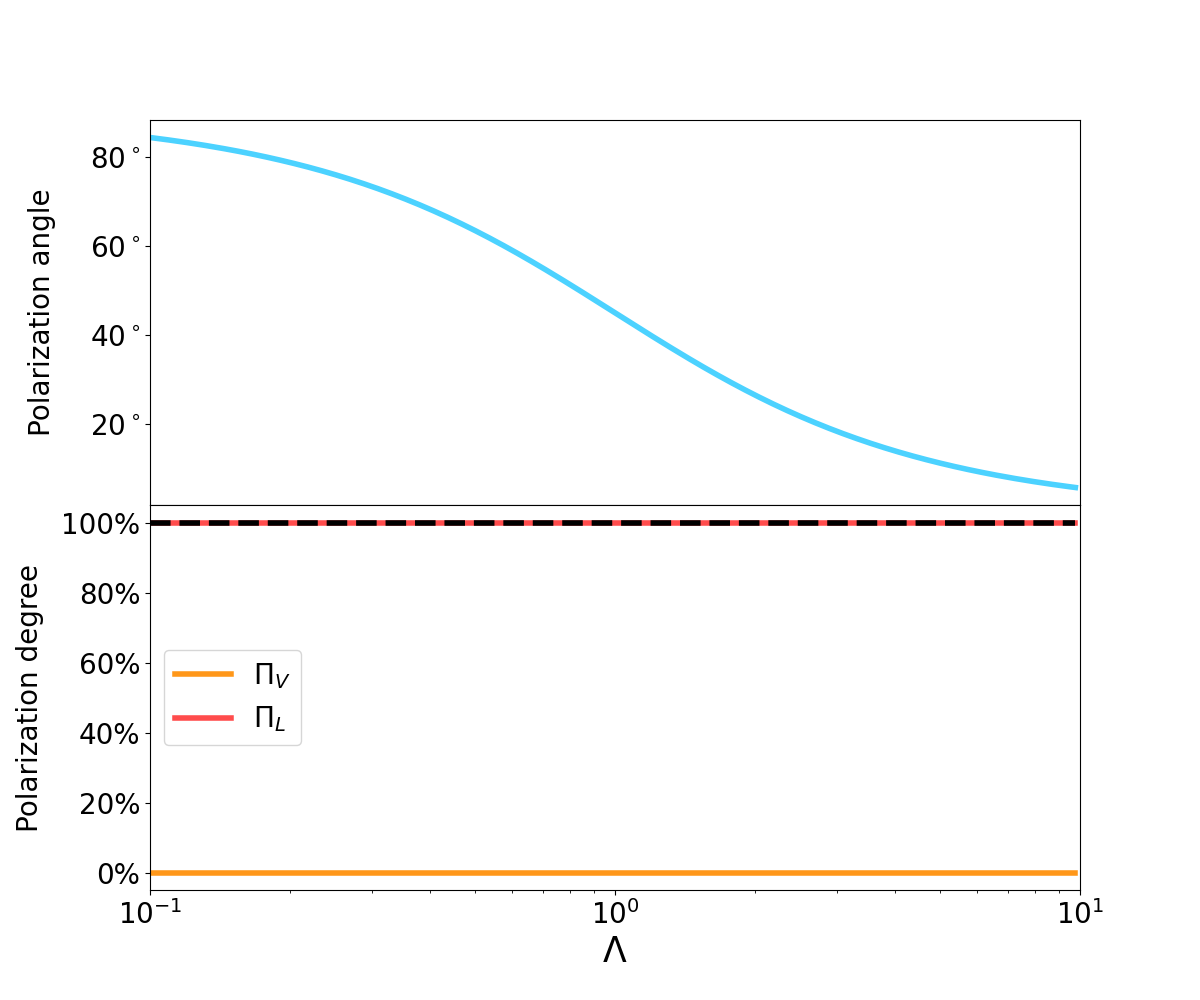}}&
\resizebox{80mm}{!}{\includegraphics[]{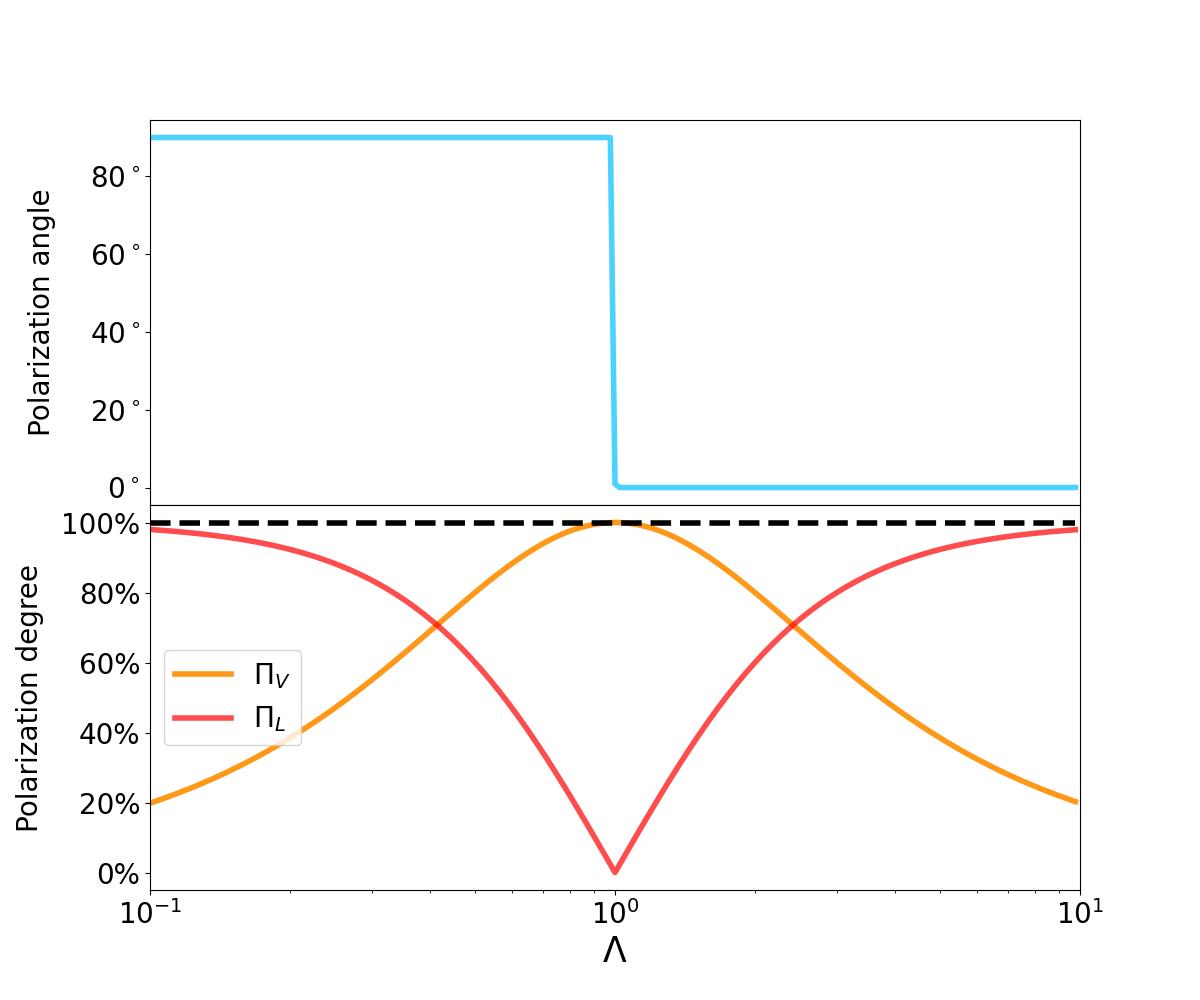}}\\
\resizebox{80mm}{!}{\includegraphics[]{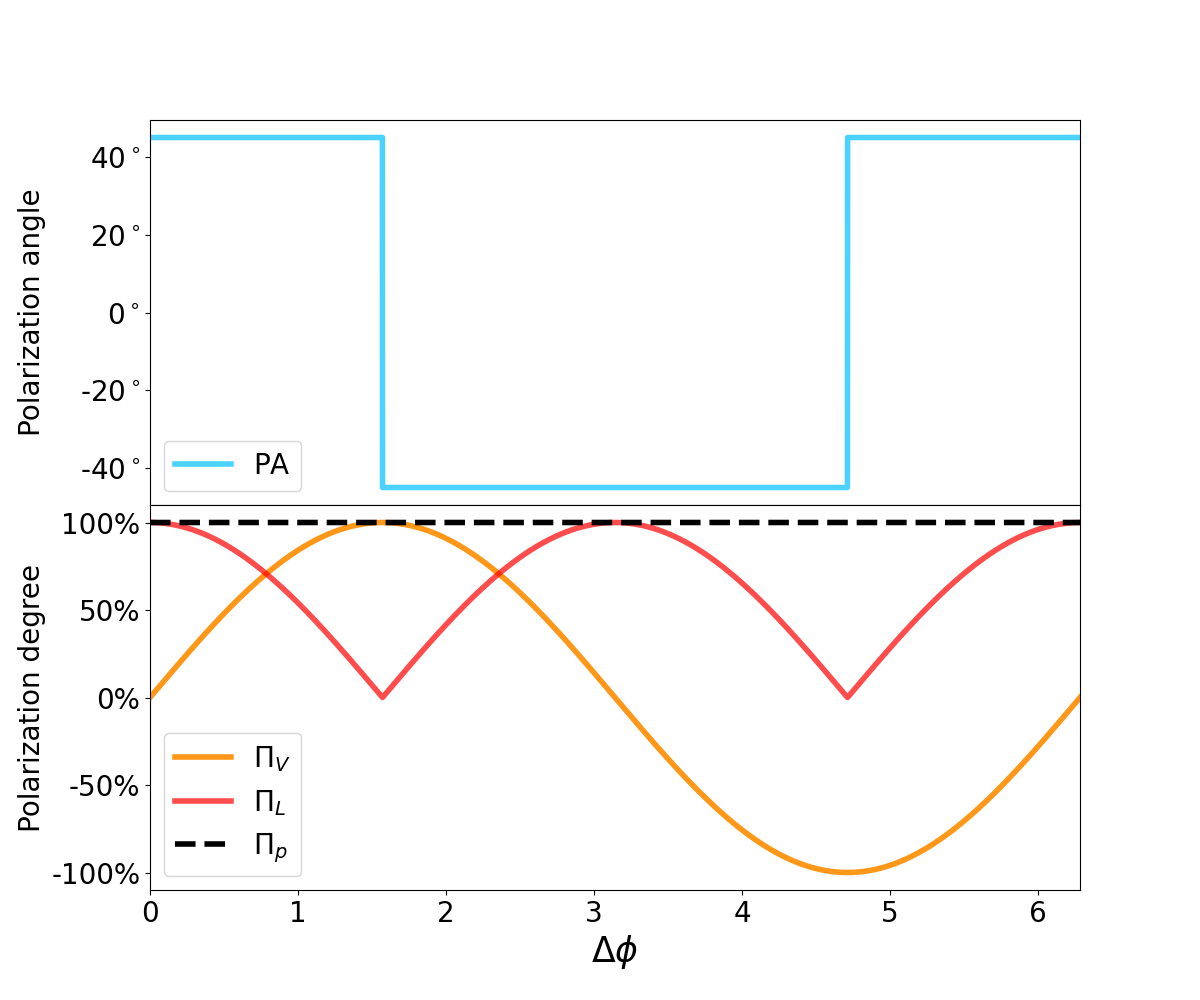}}&
\resizebox{80mm}{!}{\includegraphics[]{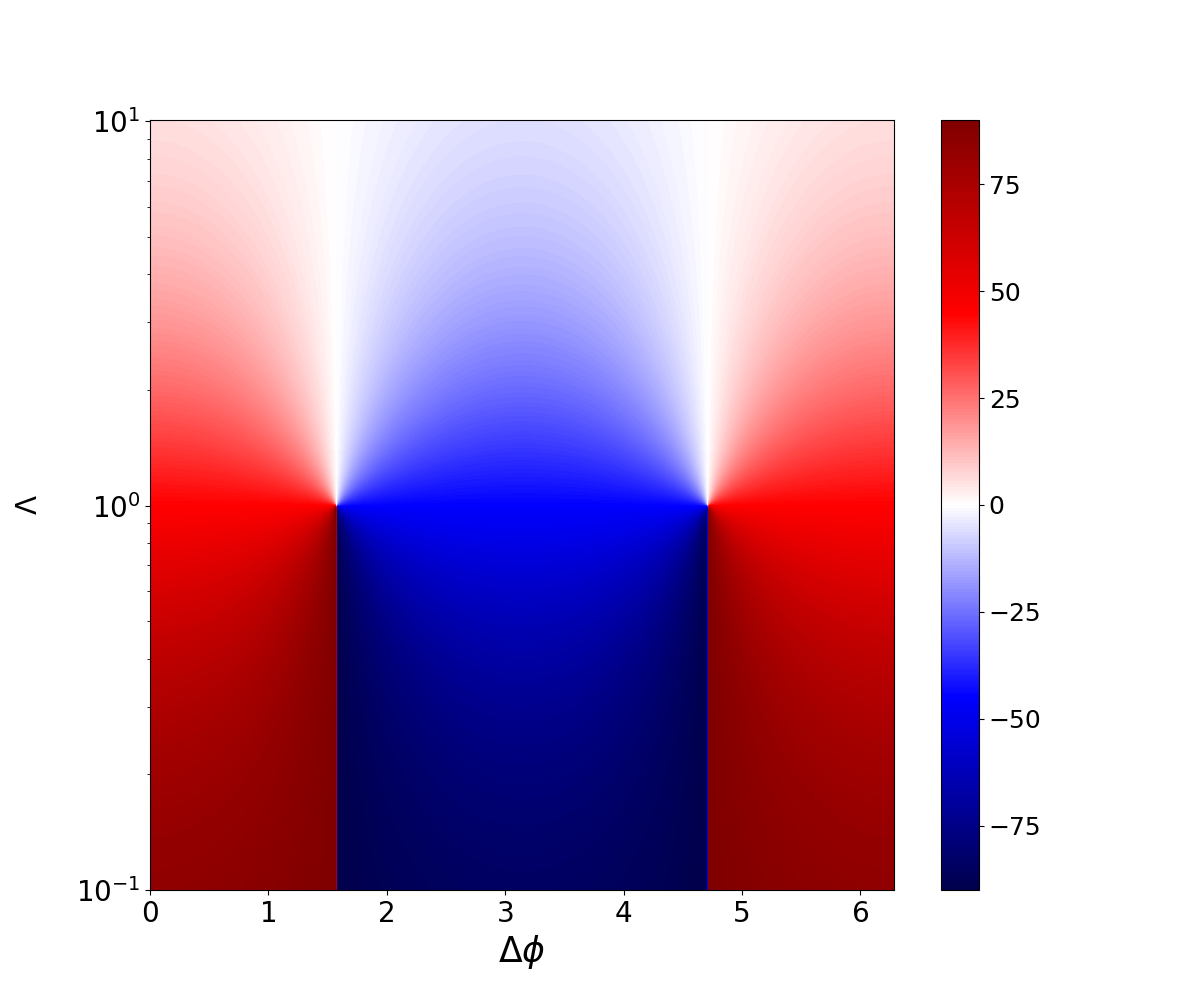}}
\end{tabular}
\caption{The polarization properties of the superposed waves after coherent superposition. 
Upper panel: PA and polarization degree as a function of amplitude ratio $\Lambda$ for $\Delta\phi=0$ (upper left) and $\Delta\phi=\pi/2$ (upper right). Two waves are initially 100\% linear polarized. 
Lower left: PA and polarization degree as a function of relative phase $\Delta\phi$ with a fixed value of $\Lambda=1$. 
Lower right: The value of PA as a function of $\Lambda$ and $\Delta\phi$.}
\label{fig:coherent_sup}
\end{center}
\end{figure*}

Coherent superposition means the wave trains are overlapped along the LOS, thus one can linearly add the electric fields and then calculate the Stokes parameters.
Thus the total orthogonal electric fields along $x$-axis and $y$-axis can be written as
\begin{equation}
E_1=E_{1x}+E_{2x}=\varepsilon_1 \cos\theta_1 e^{i\phi_{1x}}+\varepsilon_2 \cos\theta_2 e^{i\phi_{2x}}.
\end{equation}
\begin{equation}
E_2=E_{1y}+E_{2y}=\varepsilon_1 \sin\theta_1 e^{i\phi_{1y}}+\varepsilon_2 \sin\theta_2 e^{i\phi_{2y}}.
\end{equation}
The general expression of four Stokes parameters can be calculated as
\begin{equation}
I=\frac{1}{2}(E_{1\rm Re}^2+E_{1\rm Im}^2+E_{2\rm Re}^2+E_{2\rm Im}^2),
\end{equation}
\begin{equation}
Q=\frac{1}{2}[E_{1\rm Re}^2+E_{1\rm Im}^2-(E_{2\rm Re}^2+E_{2\rm Im}^2)],
\end{equation}
\begin{equation}
U=E_{\rm 1Re}E_{\rm 2Re}+E_{\rm 1Im}E_{\rm 2Im},
\end{equation}
\begin{equation}
V=E_{\rm 1Re}E_{\rm 2Im}-E_{\rm 1Im}E_{\rm 2Re},
\end{equation}
where 
\begin{equation}
E_{1\rm Re}=\varepsilon_1\cos\theta_1\cos\phi_{1x}+\varepsilon_2\cos\theta_2\cos\phi_{2x},
\end{equation}
\begin{equation}
E_{1\rm Im}=\varepsilon_1\cos\theta_1\sin\phi_{1x}+\varepsilon_2\cos\theta_2\sin\phi_{2x},
\end{equation}
\begin{equation}
E_{2\rm Re}=\varepsilon_1\sin\theta_1\cos\phi_{1y}+\varepsilon_2\sin\theta_2\cos\phi_{2y},
\end{equation}
\begin{equation}
E_{2\rm Im}=\varepsilon_1\sin\theta_1\sin\phi_{1y}+\varepsilon_2\sin\theta_2\sin\phi_{2y}.
\end{equation}

For simplicity, we take $\theta_1=0$ and $\theta_2=\pi/2$ to consider two 100\% linear polarized waves and the two electric fields are perpendicular to each other, i.e. they are orthogonal modes. 
The superposed wave polarization depends on the relative phase $\Delta\phi$ of the two waves.
The corresponding Stokes parameters can be calculated as
\begin{equation}
\begin{aligned}
&I=\frac{1}{2}(\varepsilon_1^2+\varepsilon_2^2), \ &&Q=\frac{1}{2}(\varepsilon_1^2-\varepsilon_2^2), \\
&U=\varepsilon_1\varepsilon_2 \cos\Delta\phi,  \ &&V=\varepsilon_1\varepsilon_2 \sin\Delta\phi.
\end{aligned}
\end{equation}
where $\Delta\phi=\phi_{2y}-\phi_{1x}$ is the phase difference between the two waves. 
The linear and circular polarization degrees can be calculated as
\begin{equation}
\begin{aligned}
\Pi_L&=\frac{1}{\Lambda^2+1}\sqrt{(\Lambda^2-1)^2+4\Lambda^2\cos^2\Delta\phi},
\end{aligned}
\end{equation}
and 
\begin{equation}
\Pi_V=\frac{2\Lambda}{\Lambda^2+1}\sin\Delta\phi,
\end{equation}
which is consistent with the results in \cite{Dyks2021},
where $\Lambda=\varepsilon_1/\varepsilon_2$ is the ratio of the two waves amplitude.
It should be pointed out that the total polarization degree is always constant for the coherent superposition case, i.e. $\Pi_p=\sqrt{\Pi_L^2+\Pi_V^2}=100\%$.
The PA can be expressed as
\begin{equation}
{\rm PA}=\frac{1}{2}\arctan\frac{2\Lambda\cos\Delta\phi}{\Lambda^2-1},
\end{equation}
which is consistent with the results in \cite{McKinnon2024}.
One can see that when the two waves amplitude is comparable, i.e. $\Lambda=1$, and the relative phase is $\Delta\phi=\pi/2$, 
then the Stokes parameter $Q=0$ and the sign of $U/Q$ changes, i.e. PA 90-degree jump can occur. Most importantly, $\Pi_L=0$ and $\Pi_V=100\%$.

Figure~\ref{fig:coherent_sup} shows some results of coherent superposition. In the upper left panel, we consider two 100\% linearly polarized waves superposed with the same phase ($\Delta\phi=0$) but different amplitude ratio $\Lambda$. 
As expected, no PA jump  occurs. 
PA swing can be observed with varying $\Lambda$ as the relative amplitude of the two modes varies.  The PA can vary by $90^\circ$ but no abrupt jump is seen. 
In the upper right panel of Figure~\ref{fig:coherent_sup}, we still consider two 100\% linear polarized waves but with a phase difference of $\Delta\phi=\pi/2$. 
In the time domain, an apparent abrupt PA jump is seen when $\Lambda$ crosses unity.
One can see that a $90^\circ$ PA jump is observed at $\Lambda=1$. 
Accompanied with the jump, $\Pi_L$ reaches minimum and $\Pi_V$ reaches maximum so that the total polarization fraction $\Pi_p$ remains constant.

Fixing $\Lambda=1$, we present the PA and polarization degrees as a function of phase difference $\Delta\phi$ in the lower left panel of Figure~\ref{fig:coherent_sup}. 
We obtain the same conclusion that the PA jump can occur when the two waves have same amplitude ($\Lambda=1$) but with phase difference $\Delta\phi=\pi/2$ or $3\pi/2$, at which the superposed waves have a nearly 100\% circular polarization. 
It should be pointed out that when one wave amplitude is much greater or smaller than the other wave ($\Lambda\gg1$ or $\Lambda\ll1$), 
PA jumps cannot occur and the polarization mode is nearly $\sim100\%$ linear. 
PA as a function of both $\Lambda$ and $\Delta\phi$ is presented in the lower right panel of Figure~\ref{fig:coherent_sup}. 
Again, one can see that the superposed waves are 100\% circular polarized at $\Delta\phi=\pi/2$ and $\Delta\phi=3\pi/2$ when $\Lambda=1$, accompanied by a $90^\circ$ PA jump.
Our conclusions on PA jumps are also valid for other arbitrary values of $\theta_1$ and $\theta_2$.

\subsection{Incoherent superposition}

\begin{figure*}
\begin{center}
\begin{tabular}{ll}
\resizebox{80mm}{!}{\includegraphics[]{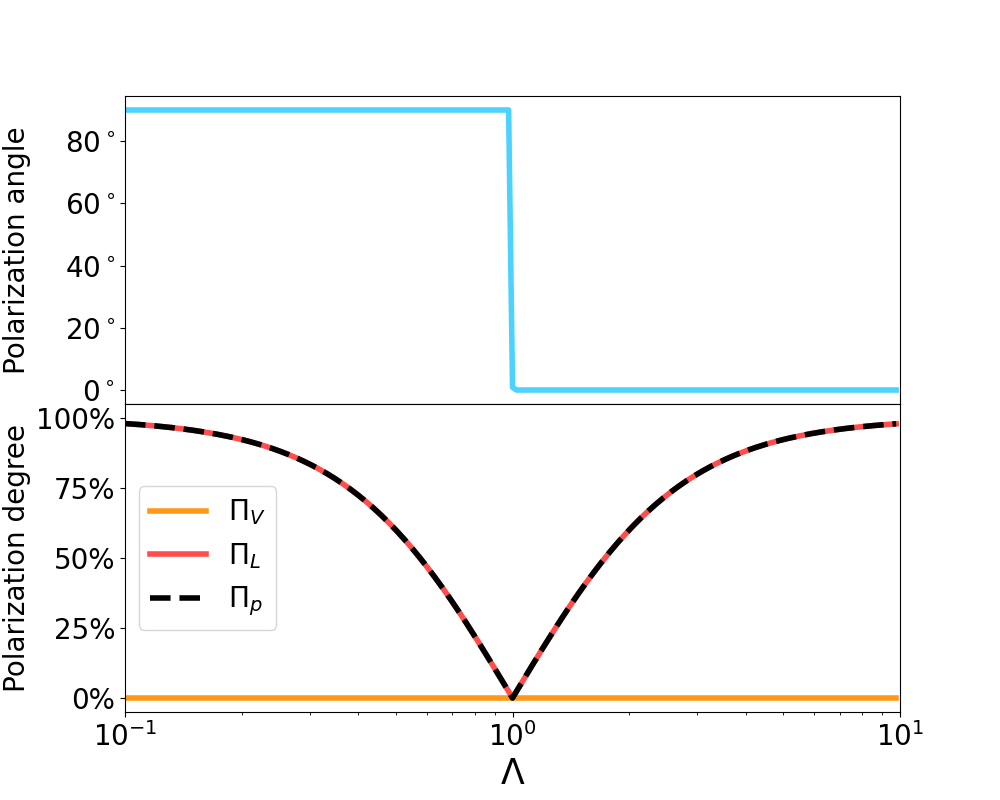}}&
\resizebox{80mm}{!}{\includegraphics[]{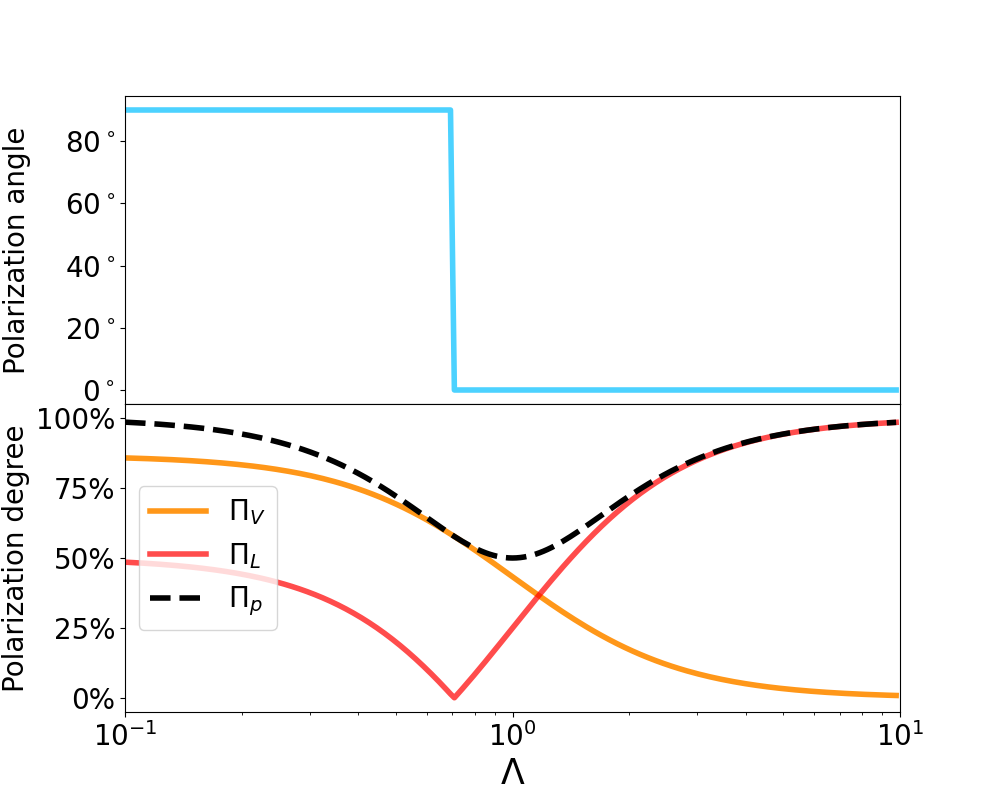}}\\
\resizebox{80mm}{!}{\includegraphics[]{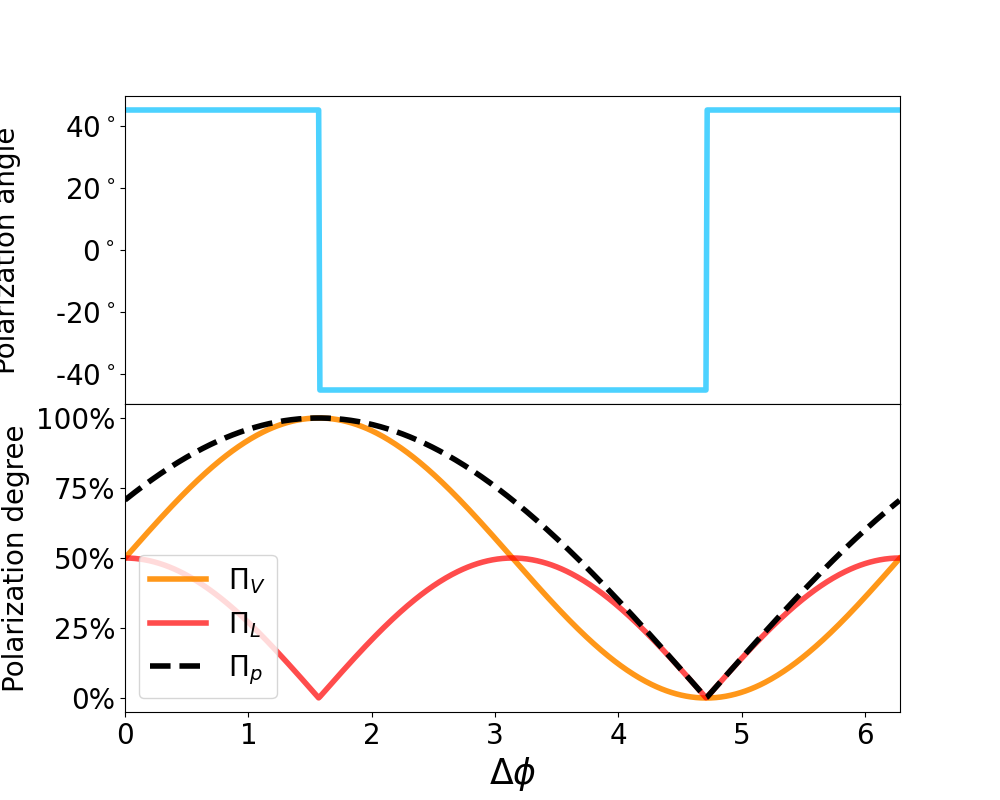}}&
\resizebox{80mm}{!}{\includegraphics[]{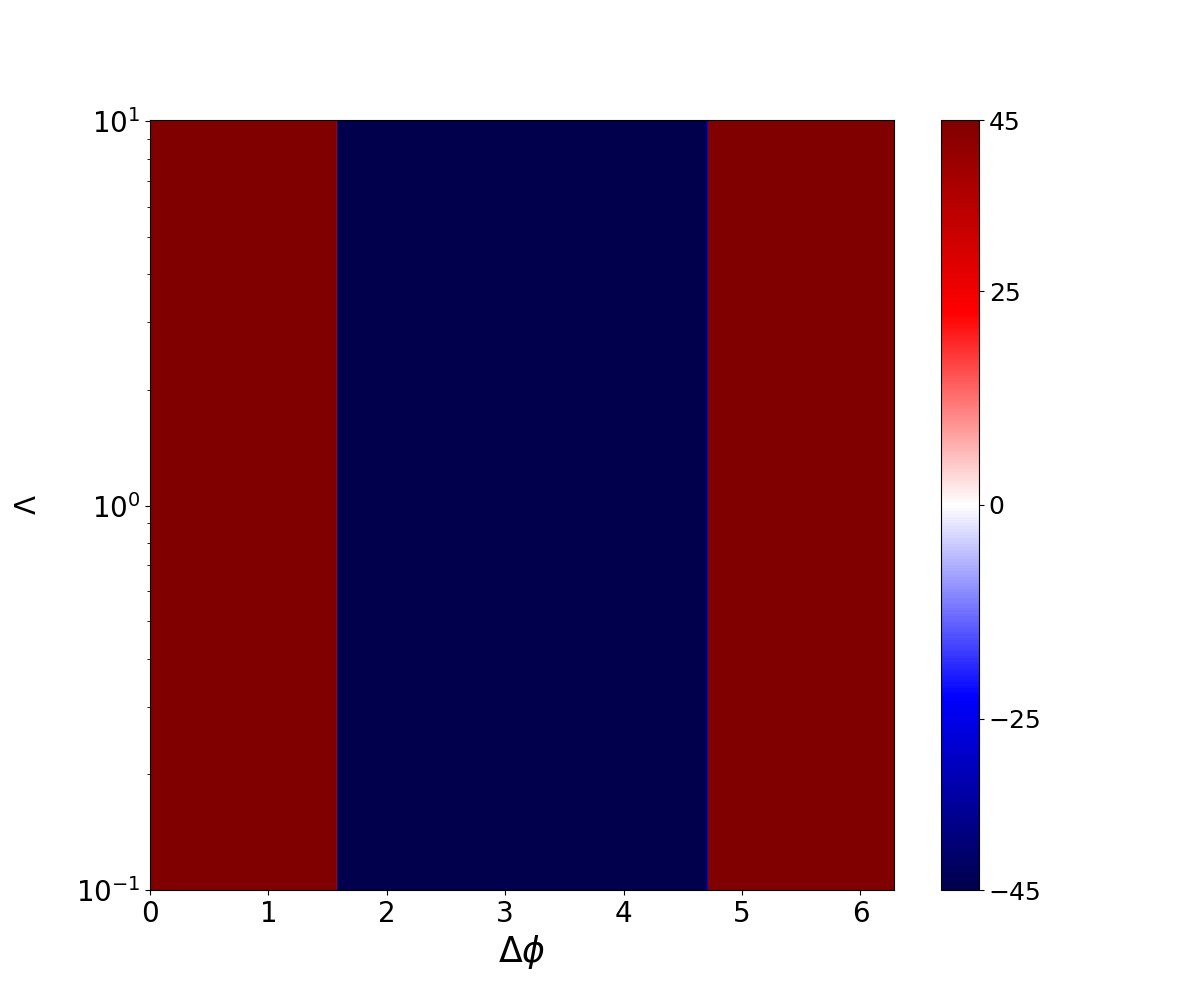}}
\end{tabular}
\caption{The polarization properties of the superposed waves after incoherent superposition. Upper panel: PA and polarization degree as a function of amplitude ratio $\Lambda$ for $\theta_1=0$, $\theta_2=\pi/2$ and $\Delta\phi_1=\Delta\phi_2=0$ (upper left). $\theta_1=0$, $\theta_2=\pi/3$, $\Delta\phi_1=0$ and $\Delta\phi_2=\pi/2$ (upper right). 
Lower left: PA and polarization degree as a function of relative phase $\Delta\phi$ with a fixed value of $\Lambda=1$, $\theta_1=\theta_2=\pi/4$ and $\Delta\phi_1=\pi/2$ (lower left). The value of PA as a function of $\Lambda$ and $\Delta\phi$ for $\theta_1=\theta_2=\pi/4$ and $\Delta\phi_1=\pi/2$ (lower right).}
\label{fig:incoherent}
\end{center}
\end{figure*}

Incoherent superposition occurs when the radiations along different LOSs are produced from very different emission regions or separated due to propagation effects. The observer detects the two waves independently, so that the Stokes parameters rather than the electric fields are summed up linearly.
For the first wave (Equation~(\ref{eq:first wave})), the Stokes parameters can be written as
\begin{equation}
\begin{aligned}
&I_1=\frac{1}{2}\varepsilon_1^2,  &Q_1=\frac{1}{2}\varepsilon_1^2(\cos^2\theta_1-\sin^2\theta_1),\\
&U_1=\varepsilon_1^2\sin\theta_1\cos\theta_1\cos\Delta\phi_1, &V_1=\varepsilon_1^2\sin\theta_1\cos\theta_1\sin\Delta\phi_1,
\end{aligned}
\end{equation}
where $\Delta\phi_1=\phi_{1y}-\phi_{1x}$.
For the second wave (Equation~(\ref{eq:second wave})), the Stokes parameters can be written as
\begin{equation}
\begin{aligned}
&I_2=\frac{1}{2}\varepsilon_2^2,  &Q_2=\frac{1}{2}\varepsilon_2^2(\cos^2\theta_2-\sin^2\theta_2),\\
&U_2=\varepsilon_2^2\sin\theta_2\cos\theta_2\cos\Delta\phi_2, &V_2=\varepsilon_2^2\sin\theta_2\cos\theta_2\sin\Delta\phi_2,
\end{aligned}
\end{equation}
where $\Delta\phi_2=\phi_{2y}-\phi_{2x}$.
Therefore, the Stokes parameters of the waves due to incoherent superposition can be calculated as
\begin{equation}
I=I_1+I_2=\frac{1}{2}\varepsilon_1^2+\frac{1}{2}\varepsilon_2^2.
\end{equation}
\begin{equation}
Q=Q_1+Q_2=\frac{1}{2}\varepsilon_1^2(\cos^2\theta_1-\sin^2\theta_1)+\frac{1}{2}\varepsilon_2^2(\cos^2\theta_2-\sin^2\theta_2).
\end{equation}
\begin{equation}
U=U_1+U_2=\varepsilon_1^2\sin\theta_1\cos\theta_1\cos\Delta\phi_1+\varepsilon_2^2\sin\theta_2\cos\theta_2\cos\Delta\phi_2.
\end{equation}
\begin{equation}
V=V_1+V_2=\varepsilon_1^2\sin\theta_1\cos\theta_1\sin\Delta\phi_1+\varepsilon_2^2\sin\theta_2\cos\theta_2\sin\Delta\phi_2.
\end{equation}
The linear and circular polarization degree can be calculated as
\begin{equation}
\Pi_L=\frac{2}{\varepsilon_1^2+\varepsilon_2^2}\sqrt{Q^2+U^2},
\end{equation}
and
\begin{equation}
\begin{aligned}
\Pi_V=\frac{2}{\Lambda^2+1}[\Lambda^2\sin\theta_1\cos\theta_1\sin\Delta\phi_1+\sin\theta_2\cos\theta_2\sin\Delta\phi_2],
\end{aligned}
\end{equation}
and the PA can be calculated as
\begin{equation}
\begin{aligned}
{\rm PA}=\frac{1}{2}\arctan\frac{2\Lambda^2\sin\theta_1\cos\theta_1\cos\Delta\phi_1+2\sin\theta_2\cos\theta_2\cos\Delta\phi_2}{\Lambda^2(\cos^2\theta_1-\sin^2\theta_1)+\cos^2\theta_2-\sin^2\theta_2}.
\end{aligned}
\end{equation}

We first consider the superposition of two 100\% linearly polarized waves, i.e. $\Delta\phi_1=\Delta\phi_2=0$. 
When the two waves have the same polarization angle, i.e. $\theta_1=\theta_2=0$, the superposed wave is still 100\% linearly polarized and there is no PA jump. 
In order to produce a PA jump, we assume that the two waves are orthogonal modes before superposition, i.e. $\theta_1=0$ and $\theta_2=\pi/2$, 
and two 100\% linearly polarized waves are assumed with $\Delta\phi_1=\Delta\phi_2=0$. 
We present PA and polarization degrees as a function of $\Lambda$ in the upper left panel of Figure~\ref{fig:incoherent}. 
One can see that the PA jump occurs when $\Lambda=1$ since the Stokes-$Q$ equals zero at the jump point and the sign also changes.

Next, we explore the superposition of two waves with different polarization properties.
We note that under certain conditions (see Section~\ref{sec:physical scenarios}), an incoherent superposition of an elliptically polarized wave and a linearly polarized wave might be possible.
We assume that the first wave is 100\% linearly polarized with $\theta_1=0$ and $\Delta\phi_1=0$, and that the second wave is elliptically polarized with $\theta_2=\pi/3$ and $\Delta\phi_2=\pi/2$. 
We present PA and polarization degrees as a function of $\Lambda$ in the upper right panel of Figure~\ref{fig:incoherent}. 
One can see that the PA jump occurs not exactly at $\Lambda=1$ which is different from the coherent superposition case. 
The linear polarization degree reaches the minimum value and the total polarization degree is not constant. 
In the lower left panel of Figure~\ref{fig:incoherent}, one can see that PA jumps occur at $\Delta\phi=\pi/2$ and the linear polarization degree reaches the minimum value for $\Lambda=1$. 
In the lower right panel of Figure~\ref{fig:incoherent}, we present PA as a function of both $\Lambda$ and $\Delta\phi$. One can see that the value of PA is constant for a constant $\Delta\phi$ regardless of $\Lambda$ and jumps can occur at $\Delta\phi=\pi/2$ or $\Delta\phi=3\pi/2$.

\

To summarize the main results of this section, we find that the polarization properties due to two waves superposition can potentially account for the observed PA jumps under certain conditions. 
Both coherent and incoherent superposition can generate PA jumps, with the linear polarization degree always reaching the minimum value. For the coherent superposition case, the total polarization fraction remains constant, so that the circular polarization reaches the maximum value when a PA jump occurs. 
On the other hand, for the incoherent superposition case, the total polarization degree does not always maintain 100\%. 
Thus, the maximum value of the circular polarization degree is not exactly at the PA jump time.
In particular, incoherent superpositions of non-orthogonal components (see upper left panel of Figure \ref{fig:incoherent}) can also produce PA jumps, although the total polarization fraction is not constant.

PA jumps can be caused by a change in the relative amplitude of the polarization modes or in the phase offset of the modes.
Physically, it seems more plausible that the observed PA jumps occur when the line of sight crosses regions where the relative intensities of the two orthogonal polarization modes vary, so that one mode gradually becomes dominant over the other. 
In contrast, producing PA jumps solely through the phase offset between the modes evolving to precisely $\pi/2$ or $3\pi/2$ would likely require a more fine-tuned plasma or propagation condition and therefore seems less generic.

\section{GENERIC OBSERVATIONAL CONSTRAINTS}\label{sec:general constraint}

\begin{figure*}
    \includegraphics[width=18 cm,height=8 cm]{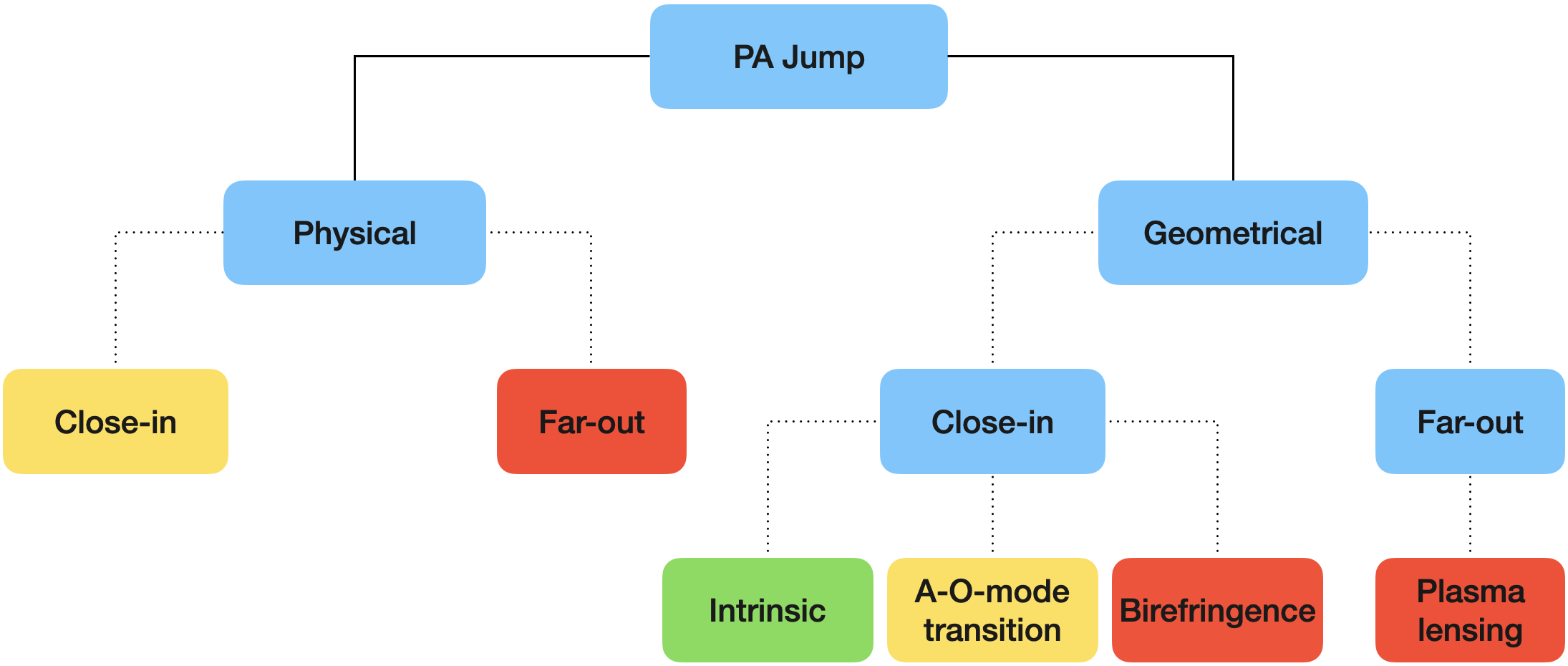}
    \caption{General physical and geometrical processes to generate polarization angle jumps of FRBs discussed in Section~\ref{sec:general constraint}. 
    For each process, close-in (inside the magnetosphere) and far-away (outside the magnetosphere) models are investigated.
    The favored processes are marked as green, possible process is marked as yellow and the disfavored processes are marked as red.}
    \label{fig:map}
\end{figure*}

So far, we have not discussed PA variation/jump as a function of time, as is observed in the three FRB bursts (and also in many radio pulsars). In order to account for the observed jumps within the milliseconds timescale, some very generic constraints on the models can be placed. This is the purpose of this section. Various possible mechanisms to produce PA orthogonal jumps and their plausibility are presented in Figure~\ref{fig:map}.

\subsection{Physical vs. Geometrical Scenarios}

To observe a sudden change of PA within the milliseconds timescale, one can envisage two distinct scenarios. The first scenario is that there is an intrinsic change of the physical properties of the emitter within such a short timescale, which we call the physical scenario. 
Another is to introduce a geometric effect, 
the object's rotation sweeps its emission region across the LOS,
and the sudden jump in PA occurs as the LOS goes across a region where the physical conditions change. We call this second scenario the geometric scenario. 

We argue that the physical scenario is less plausible, even though physical processes may, in principle, give rise to such an abrupt change under special conditions. Without a geometric effect, the emission along the LOS is likely dominated by one mode.
If the emission region of an FRB is not stochastically varying, and if the emission is dominated by one mode during millisecond timescale, 
in order for the PA of the emission mode to suddenly jump by $90^\circ$ within milliseconds, one requires that the magnetic configuration in the emission region undergo an abrupt change during such a short period of time. Such a scenario has never been realized and even envisaged in FRB emission models involving a typical magnetar with a spin period of the order of seconds\footnote{A millisecond magnetar can have abrupt change of the plasma properties within milliseconds, but these sources have a spindown timescale of thousands of seconds, too short to interpret actively repeating FRBs that sustain bursting emission for months to years.}, invoking either magnetospheric emission or maser emission in relativistic shocks. The ms-duration also places a tight limit of the size of the emission region, which is
\begin{equation}\label{eq:lightcrossing}
l_{\rm em}=c\delta t=(3\times10^{7} \ {\rm cm}) \ \delta t_{-3},
\end{equation}
for a non-relativistic emitter, 
where $\delta t$ is the typical duration of the jump (which is normalized to millisecond). Furthermore, even if the magnetic field undergoes a global reconfiguration in such a short period of time, it is very contrived to allow the PA jump by $90^\circ$. One may also envisage a sudden change of the particle spatial distribution or radiation mechanism to account for the jump without a significant change of the magnetic configuration. However, the required change is also highly contrived and there is no known physics to trigger such a change.

It should be pointed out that single pulse studies of radio pulsars reveal that the PA can exhibit variability and discontinuity on timescales as short as microseconds \citep{Cordes1976,Cordes&Hankins1977,Singh2024} or even nanoseconds \citep{Hankins2003}.
Moreover, PA jumps can occur across the entire pulse rather than in a fixed phase, and PA discontinuities may appear at essentially arbitrary pulse phases \citep{Manchester1975,Cordes1978,Backer&Rankin1980,Stinebring1984}.
These observations indicate that the magnetospheric environment of neutron stars can be highly dynamic, evolving rapidly on very small temporal and spatial scales.
Such short variability suggests that under certain circumstances, plasma processes or dynamic background magnetic field configurations, could in principle induce PA changes. However, since FRBs have brightness temperatures over 10 orders of magnitude higher than pulsars, it is unclear whether the stochastic pulsar mechanism can be straightforwardly applied to FRBs before observationally the stochastic nature of the FRB emission is verified.

Although the phenomenology of pulsar radio emission implies that physical processes inside the magnetosphere cannot be entirely ruled out (marked as yellow in Figure~\ref{fig:map}) and far-out effects are unlikely (marked as red), 
we conclude that the origin of the PA orthogonal jump very likely invokes a geometric scenario, with a rotating object viewed by a varying viewing angle and different emission modes dominating in different viewing directions.

\subsection{Close-in vs. Far-out}

Within the geometric scenario, one can generally discuss two types models. The first type invokes magnetospheric emission from an FRB central engine, which is called close-in or pulsar-like models; the second type invokes relativistic shocks way outside of the light cylinder of the engine, which is called far-out or GRB-like models. We argue that the observational data strongly disfavors the far-out models.

First, within the far-out models, because the emission region is far from the central engine and well beyond the light cylinder, the emitter is not undergoing significant rotation. 
The emission region does not move significantly across the LOS within milliseconds.
More specifically, the synchrotron maser model \citep{Plotnikov&Sironi2019,Metzger2019,Sironi2021} can produce both X-mode and O-mode emissions, but the wave amplitude ratio between the two modes is found in the 3D PIC simulation to be dominated by the X-mode, i.e. $|{E_{\rm X}}/{E_{\rm O}}|\sim1.6\sqrt{\sigma}\gg1$ \citep{Sironi2021},
where $\sigma \gg 1$ is the magnetization factor, which is required to be much greater than unity to allow the mechanism to produce the high brightness temperature as observed in FRBs. 
As a result, even if the central engine rotation can bring different emission regions to the LOS, the observed emission is consistently dominated by the X-mode.
Without a contrived sudden spatial change of the magnetic configuration (which is never observed in numerical simulations), the sudden orthogonal jump is impossible.

Another possibility for producing an orthogonal jump from far away distances is through plasma lensing. In particular, it has been shown \citep{Er2023} that with a proper setup, a plasma lens can produce distinct cusps across which a jump in the polarization mode can happen. 
However, the probability that the rotation of the central engine brings these cusps into the LOS during an FRB burst is extremely low.
Consider the proper motion velocity of a magnetar with $v_{\rm pro}=10^7 \ \rm cm \ s^{-1}$.
Within the typical time duration of PA jumps $\delta t\sim 1 \ \rm ms$, the corresponding transverse length scale is $l_{\perp,\rm pro}\simeq v_{\rm pro}\delta t=(10^4 \ {\rm cm}) \ v_{\rm pro,7}\delta t_{-3}$.
Notice that three PA jump events were observed in over 2000 bursts (1863 bursts in the first episode and more than 600 bursts in the second episode) from FRB 20201124A are reported during the time span of $\sim 2.5$ months \citep{NiuJR2024}.
Consider the typical duration of an FRB $\Delta t_{\rm frb}$,
the total time duration of $\sim 2000$ FRBs may be estimated as $\Delta T\sim 2000 \Delta t_{\rm frb}=(2 \ {\rm s}) \ \Delta t_{\rm frb,-3}$, which is related to the distance traveled by the proper motion of the FRB engine\footnote{The true distance traveled during the 2.5-month timescale is much longer than this, but most of it is ``unobservable'' for plasma lensing because there is no radio emission detected.} 
\begin{equation}
l_{\perp,\rm obs}\simeq v_{\rm pro} \Delta T\simeq (2\times10^{7} \ {\rm cm}) \ v_{\rm pro,7}\Delta T_{0.3}.
\end{equation}
One can then derive a number density of the plasma lens normal to the line of sight
\begin{equation}\label{eq:linear n_l}
n_l\simeq\frac{3}{l_{\perp,\rm obs}}\simeq (1.5\times10^{-7} \ {\rm  cm^{-1}}) \ v_{\rm pro,7}^{-1}\Delta T_{0.3}^{-1}.
\end{equation}
This density is way too high for plasma lenses. Even if one considers that these lenses are placed one by one without any spacing (which is realistically impossible because a special arrangement of the plasma properties is needed to make a lens), the linear density is still smaller than Equation~(\ref{eq:linear n_l}). This can be proved by estimating the Fresnel angle ($\theta_{\rm F}$), which is defined as the angular position on the lens plane where the radio wave from the source propagating through this point would induce an additional geometric phase difference of $2\pi$ relative to a straight line trajectory from the source to the observer. 
The Fresnel scale can be estimated as 
\begin{equation}
\begin{aligned}
d_{\rm F}=\left(\frac{2\lambda d_{\rm SL}d_{\rm LO}}{d_{\rm SO}}\right)^{1/2}&\simeq \left({2\lambda d_{\rm SL}}\right)^{1/2}\\
&\simeq(2.4\times10^7 \ {\rm cm}) \ \nu_9^{1/2}d_{\rm SL,13}^{1/2},
\end{aligned}
\end{equation}
where $d_{\rm LO}$ is from the lens to the observer, $d_{\rm SO}$ is from the FRB source to the observer, and $d_{\rm SL}$ is the distance from the FRB source to the lens\footnote{The non-linear parameter of FRBs at the magnetar wind zone can be estimated as $a\simeq1.6\times10^4 \ L_{\rm frb,42}^{1/2}\nu_9^{-1}r_{13}^{-1}$. 
Particles located ahead of the FRB wavefront are rapidly accelerated to relativistic speeds, which effectively reduces the refractive index. For this reason, $d_{\rm SL}$ is normalized to a typical distance of $10^{13}$ cm to ensure that the non-linear parameter does not significantly exceed unity.
}.
This gives a maximum linear number density for lenses 
\begin{equation}
n_{\rm l,max} \sim n_{\rm F}\sim 1/d_{\rm F}\simeq (4.2\times10^{-8} \ {\rm cm^{-1}}) \ \nu_9^{-1/2}d_{\rm SL,13}^{-1/2},
\end{equation}
which is smaller than $n_l$ defined in Equation~(\ref{eq:linear n_l}). The real linear number density of the lens in the ISM is not well constrained, but it should be much smaller than this maximum value.
We conclude that PA jumps cannot be produced through plasma lensing outside the magnetosphere of the central engine.

\section{Physical scenarios for orthogonal jumps inside the magnetosphere}\label{sec:physical scenarios}

\begin{figure*}
    \includegraphics[width=18 cm,height=12.5 cm]{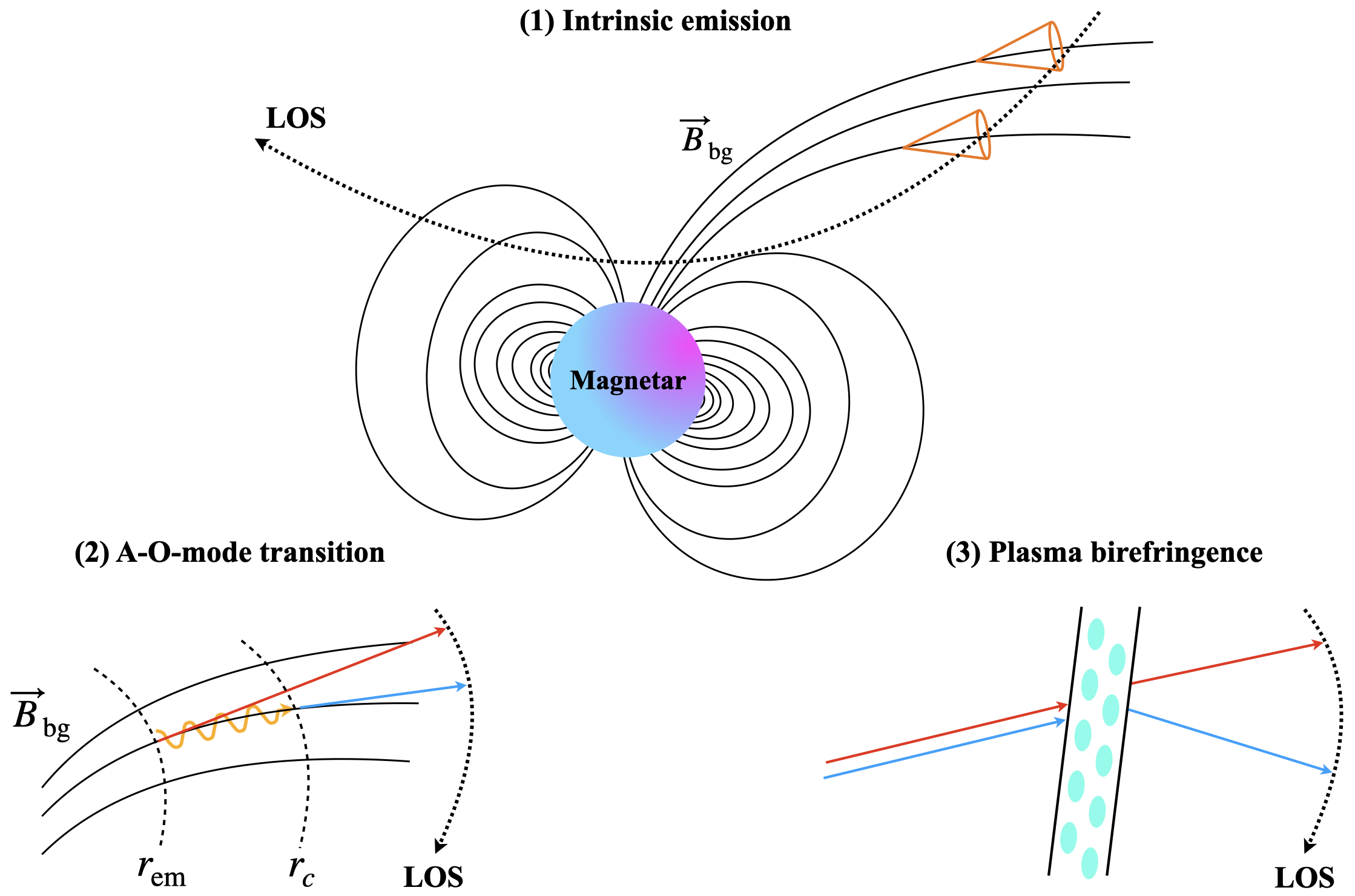}
    \caption{A cartoon figure for three scenarios to produce PA jumps of FRBs including (1) intrinsic emission (upper panel), (2) A-O-mode conversion (lower left) and (3) Plasma birefringence (lower right) discussed in geometric effects.
    In both (2) and (3),  
    the red line and blue line denote X-mode and O-mode of FRBs, respectively.
    The curved back dashed arrow represents the trajectory of the LOS, with its direction indicating how the LOS evolves over time. 
    The solid black curves denote background magnetic field ($\vec B_{\rm bg}$) in both (1) and (2).
    In (1), 
    as the source rotates azimuthally, the emission transitions from an O-mode-dominated region to an X-mode-dominated region across the fixed LOS.
    In (2), two dashed black lines denote the emission radius ($r_{\rm em}$) and critical conversion radius ($r_c$), respectively.
    In (3), the cyan clumps represent plasma located ahead of the FRBs.
    }
    \label{fig:geometric_cartoon}
\end{figure*}

The generic constraints presented in the last section leave the geometric scenarios of the magnetospheric origin as the most plausible ones for orthogonal jumps. In this section, we investigate in detail the three most relevant magnetospheric processes, as indicated in Figure~\ref{fig:geometric_cartoon}: 
\begin{enumerate}
\item The two orthogonal modes (X-mode and O-mode) are intrinsically produced\footnote{
Both X-mode and O-mode waves can escape from the magnetar magnetosphere in the open field line region, where the plasma streams relativistically along the background magnetic field \citep{QKZ,Lyutikov2024,Huang2024}. 
Thus scattering and damping processes are unlikely to impact the amplitudes of the two orthogonal modes.}. 
The relative dominance of these two modes depends on the viewing angle, the geometry and plasma properties of the emission region. 
As the rotation of the central engine successively brings different emission regions into the LOS,
PA jumps occur naturally when the dominant mode switches between the X-mode and O-mode.
\item A-O mode conversion: The emission is produced in the high-density plasma region, with the X-mode escaping freely in straight lines. The orthogonal mode is in the Alfv\'en mode which propagates along the magnetic field lines. As Alfv\'en waves finally escape in a low-density region as O-mode waves, they naturally beam toward a different direction from the X-mode waves.  
As the central engine rotates, the observer will first detect the X-mode and later the O-mode (or vice versa). A PA jump would be observed when the rotation successively brings regions dominated by the two different modes into the LOS.
\item Plasma birefringence: If a wave with mixed X and O modes reaches a plasma region with an incident angle different from $90^\circ$, the two modes would be refracted in different directions. 
A PA jump would occur when the emission regions dominated by two different modes sweep across the LOS as the star rotates.
\end{enumerate}

In this section, we assume that the X- and O-modes are orthogonal and mutually incoherent, which are consistent with the observed FRB PA jumps in FRB 20201124A \citep{NiuJR2024}.

\subsection{Intrinsic emission mechanisms}\label{sec:Intrinsic emission mechanisms}

Some intrinsic radiation mechanisms can produce two orthogonal modes (e.g. X- and O- modes). In order to detect a sudden jump between the two modes, one requires three conditions: (1) the medium should be transparent to both modes; 
(2) as the source rotates, the observer can detect two regions dominated by the two modes, respectively.
and (3) the two modes should have comparable amplitudes at the transition time.

Before discussing the detailed intrinsic mechanisms, we first consider the condition for the transparency of the two modes. 
When the wave frequency exceeds the plasma frequency, both for X-mode and O-mode, electromagnetic waves can propagate freely such as in vacuum. When the frequency is below the plasma frequency,
the X-mode is in the form of fast magnetosonic waves \footnote{The X-mode and fast magnetosonic waves have the same polarization with the electric field perpendicular to the ($\vec k - \vec B_{\rm bg}$) plane and the same dispersion relation ($\omega=kc$) within the magnetar magnetosphere \citep{Stix1992}. The difference is that the X-mode is defined above the plasma frequency whereas the fast magnetosonic mode is defined below the plasma frequency.}, which can still propagate as long as they remain in the linear regime, where the MHD description remains valid\footnote{The MHD description may break down when the amplitude of the fast magnetosonic wave becomes comparable to background magnetic field, 
which may occur in the closed field line region and when fast magnetosonic waves steepen into monster shocks \citep{ChenYR2022,Beloborodov2023}.
In the open field line region, the radius for such a nonlinear regime is greater than the conversion radius, which means that fast magnetosonic waves already convert into X-mode waves before driving a monster shock.
}. 
However, the O-mode cannot propagate, which can only be converted from the Alfv\'en mode at a critical radius $r_c$ defined by the condition $\omega'=\omega_p'$ in the comoving frame of the relativistic plasma.
The conversion radius for Alfv\'en waves (to O-mode) and for fast magnetosonic waves (to X-mode) can be calculated as
\begin{equation}\label{eq:r_c}
\begin{aligned}
r_{c}&=\left(\frac{\xi B_\star R_\star^3q {\cal D}^2}{cP\gamma m_e \pi\nu^2}\right)^{1/3}\\
&\simeq
\left\{
\begin{array}{ll}
&(1.3\times 10^8 \ {\rm cm}) \ \xi^{1/3} B_{\star,15}^{1/3}R_{\star,6}\gamma_2^{1/3}P^{-1/3}\nu_9^{-2/3}, \\
&~~~~~~~~~~~~~~~~~~~~~~{\rm A~mode \rightarrow O~mode},\\
&(5.2\times10^7 \ {\rm cm}) \ \xi^{1/3}B_{\star,15}^{1/3}R_{\star,6}\gamma_2^{-1/3}P^{-1/3}\nu_9^{-2/3}\left(\frac{{\cal D}}{50}\right)^{2/3}, \\
&~~~~~~~~~~~~~~~~~~~~~~{\rm F~mode \rightarrow \ X~mode},
\end{array}
\right.
\end{aligned}
\end{equation}
where $\xi$ is the multiplicity, 
$B_\star$ is the magnetar surface magnetic field strength, $R_\star$ is the magnetar radius,
$\nu=\omega/2\pi$ is the FRB frequency, $\gamma$ is the Lorentz factor of the relativistic plasma, ${\cal D}=1/[\gamma(1-\beta\cos\theta)]$ is the Doppler factor, which is adopted as ${\cal D} = 2 \gamma$ for Alfv\'en waves with $\theta=0$.
For fast magnetosonic waves, $\cal D$ is normalized to 50, which corresponds to $\theta=1^\circ$ for $\gamma=100$.

The second condition requires that the two emitters that have orthogonal modes are located in different regions in the magnetosphere, 
with the X- or O-mode-dominated regions sweep across the LOS as the central object rotates.
This is rare but can happen under special geometries, and when it happens, the superposition is incoherent and typically leads to significant depolarization\footnote{ 
We also note that the total polarization degree may not be strictly equal to zero, thus a more plausible scenario may be partially coherent superposition \citep{Oswald2023}.}. 
All these are consistent with the observations.

The third condition, i.e. the X- and O-modes have comparable amplitudes, depends on the concrete radiation mechanisms. In the following, we discuss the some widely discussed radiation mechanisms, curvature radiation (CR), inverse Compton scattering (ICS), and precursor of monster shock, in detail.

\subsubsection{Coherent curvature radiation}

A widely studied radiation mechanism for FRBs within the inner magnetar magnetosphere is the coherent curvature radiation emitted by charged bunches \citep{Kumar2017,Lu2020,Kumar&Bosnjak2020}. 
We consider one single charged bunch moving along the background magnetic field line, 
the amplitudes of orthogonal modes via curvature radiation are given by \citep{Jackson1998}
\begin{equation}
\begin{aligned}
E_{\rm X}(\omega)\simeq\frac{i2\rho}{\sqrt{3}c}\left(\frac{1}{\gamma^2}+\theta_v^2\right)K_{2/3}(\varsigma)
\end{aligned}
\end{equation}
which describes the X-mode amplitude,
and
\begin{equation}
E_{\rm O}(\omega)\simeq\frac{2\rho}{\sqrt{3}c}\theta_v\left(\frac{1}{\gamma^2}+\theta_v^2\right)^{1/2}K_{1/3}(\varsigma)
\end{equation}
which describes the O-mode amplitude. 
Here, $\rho$ denotes the curvature radius, $\theta_v$ is the viewing angle between the LOS and the particle motion direction, 
$K_{\nu}(\varsigma)$ is the modified Bessel function
and the parameter $\varsigma$ is defined by $\varsigma=\omega\rho(1/\gamma^2+\theta_v^2)^{3/2}/3c$.
The amplitude ratio of the two orthogonal modes can be calculated as
\begin{equation}\label{eq:curvature_ratio}
\left|\frac{E_{\rm X}}{E_{\rm O}}\right|=\frac{1}{\theta_v}\left(\frac{1}{\gamma^2}+\theta_v^2\right)^{1/2}\frac{K_{2/3}(\varsigma)}{K_{1/3}(\varsigma)}\simeq 1 \  \ {\rm for} \  \ \gamma\theta_v\gg1.
\end{equation}
Notice that the values of $K_{2/3}(\varsigma)$ and $K_{1/3}(\varsigma)$ are nearly the same.
When $\theta_v=0$ (on-axis case), the electron can only produce 100\% linearly polarized X-mode waves since $E_{\rm O}(\omega)=0$. 
In order to produce non-negligible O-mode waves, one needs to observe the emitting bunch at an off-axis viewing angle ($\theta_v\neq0$).
We note that the ratio $E_{\rm X}/E_{\rm O} > 1$ when $\theta_v < 1/\gamma$ and approaches unity when $\theta_v > 1/\gamma$, 
i.e. the amplitude ratio of X-mode to O-mode is always greater than unity.

In order to obtain the temporal evolution of PA, we need to find out the relation between the amplitude ratio and time.
The viewing angle dependence on observation time as the magnetar spins can be written as
\begin{equation}
\theta_v (t)=\frac{v_{\rm rot} t}{r_{\rm em}}+\theta_{v,0}=\frac{2\pi t}{P},
\end{equation}
where the initial viewing angle $\theta_{v,0}=0$ is chosen at $t=0$.
The amplitude ratio of two modes depends on $\theta_v$, thus it also depends on time as magnetar spins.
It should be pointed out that PA jumps do not occur solely via intrinsic wave superposition since $E_{\rm X}\gtrsim E_{\rm O}$ is always satisfied for a single charged bunch's curvature radiation.
Observations suggest an incoherent superposition.
The X-mode remains dominant at all viewing angles, even though different emitting bunches at different locations sweep across the LOS.
It is difficult to produce an orthogonal PA jump unless the magnetic configuration switches by $90^\circ$, which is difficult to achieve. 
Thus, additional propagation effects are needed to make the O-mode waves dominant.

\subsubsection{Coherent inverse Compton scattering}

\begin{figure*}
\begin{center}
\begin{tabular}{ll}
\resizebox{91mm}{!}{\includegraphics[]{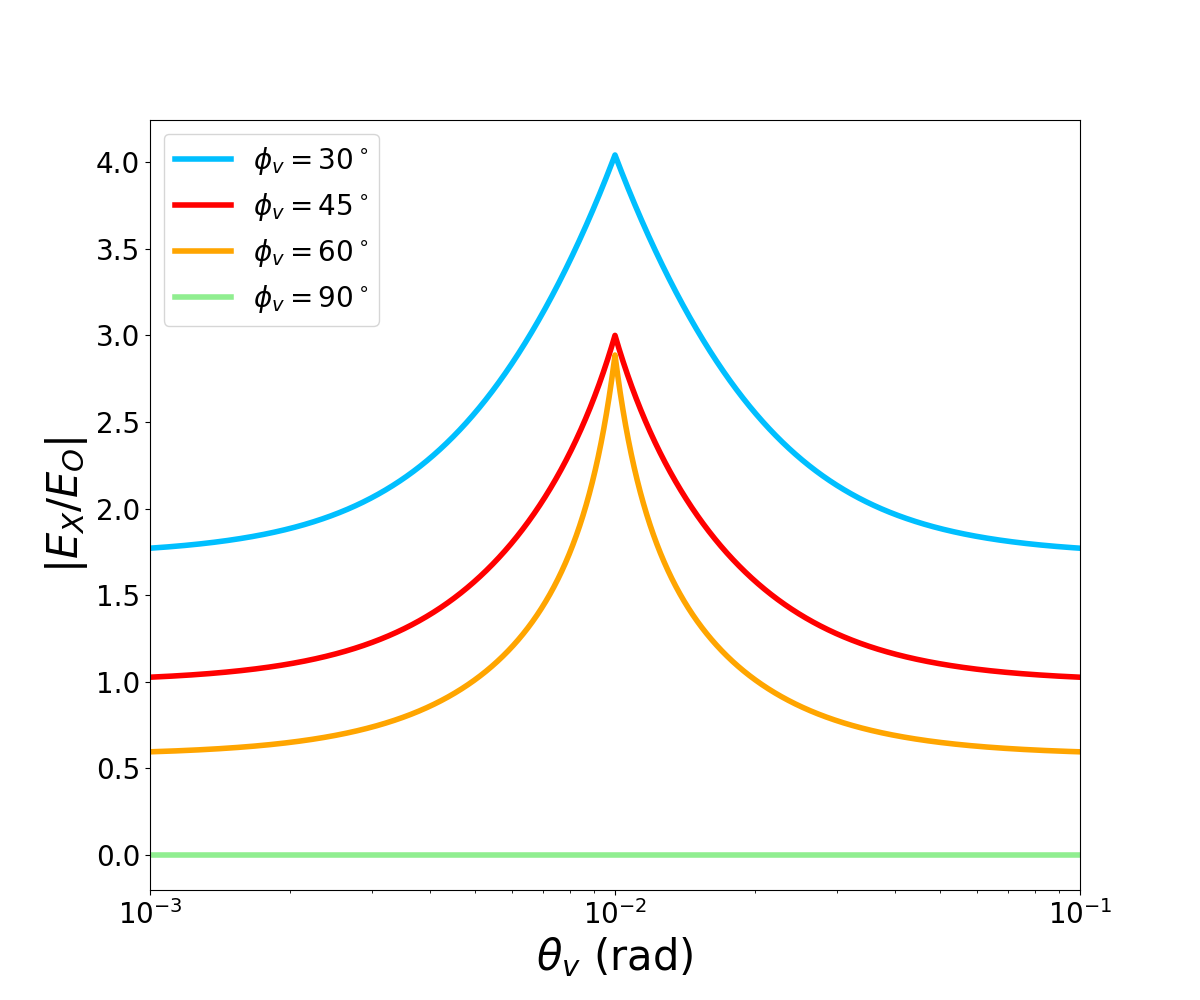}}&
\resizebox{92mm}{!}{\includegraphics[]{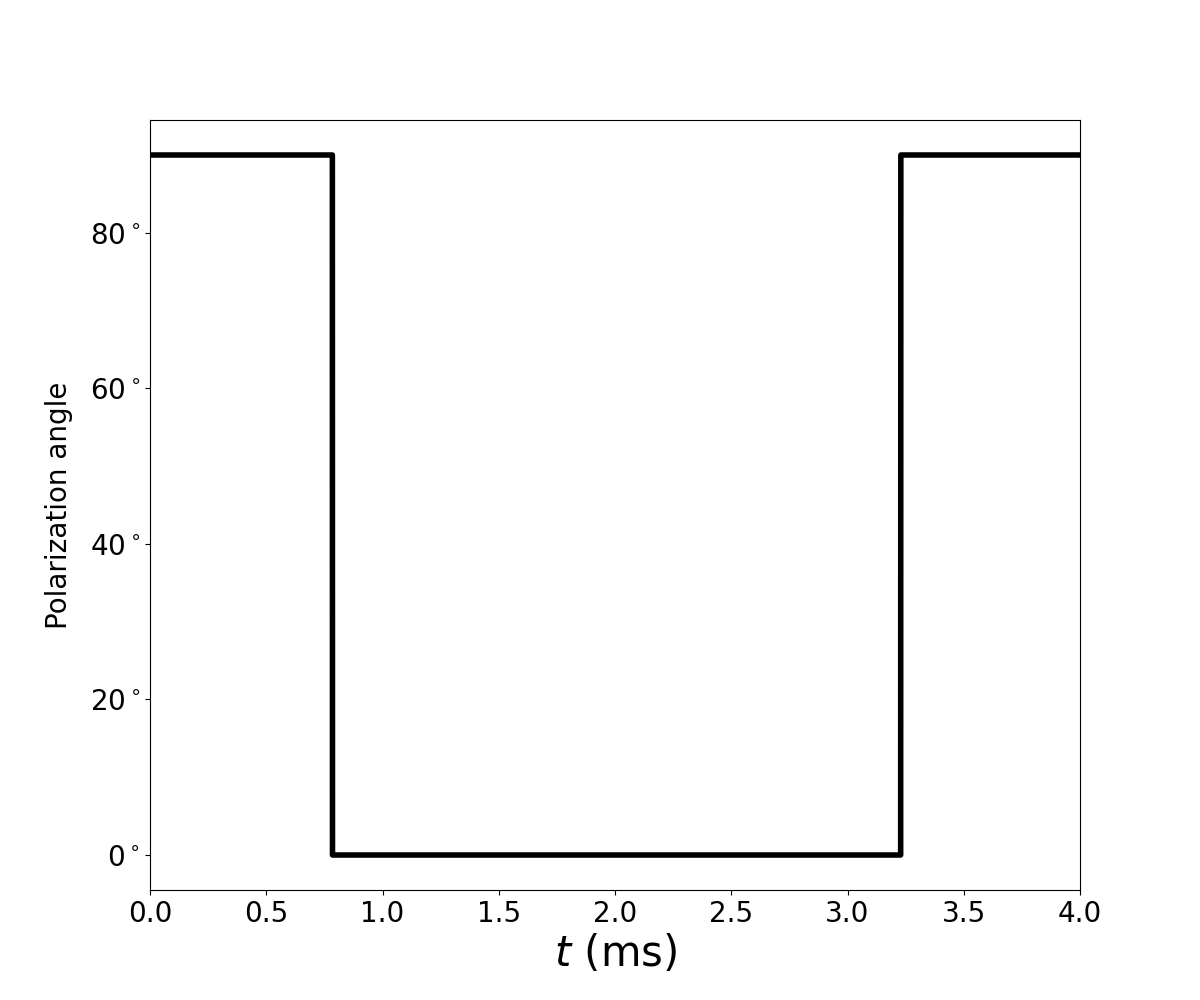}}
\end{tabular}
\caption{Left panel: The amplitude ratio of X-mode to O-mode as a function of viewing angle $\theta_v$ for ICS.
Right panel: Polarization angle as a function of time via incoherent superposition.
The Lorentz factor $\gamma=100$, magnetar spin period $P=1 \ \rm s$ and $\phi_v=60^\circ$ are adopted for the right panel.}
\label{fig:amplitude ratio}
\end{center}
\end{figure*}

The coherent ICS process by charged bunches off $\sim{\rm kHz}$ frequency fast magnetosonic waves produced by the magnetar crust quake in the context of FRBs has been discussed by \cite{Zhang22} and \cite{Qu&Zhang2024}.  
We define the unit vector of the LOS as $\hat{n}'=\sin\theta_v'\cos\phi_v'\hat{x}+\sin\theta_v'\sin\phi_v'\hat{y}+\cos\theta_v'\hat{z}$ in the comoving frame of the relativistic charged bunch.
The electric field of the scattered waves for one charged particle in the comoving frame is given by \citep{Qu&Zhang2024}
\begin{equation}\label{eq:single particle's field}
\begin{aligned}
\vec E_{\rm rad}'&=\frac{q^2E_{\rm fmw}'\sin(\omega_i't')}{m_ecR'}\left(\frac{\omega_i'}{\omega_B'}\right)[\cos\phi_v'\sin\phi_v'\sin^2\theta_v'\hat{x}\\
&-(\cos^2\theta_v'+\cos^2\phi_v'\sin^2\theta_v')\hat{y}+\cos\theta_v'\sin\phi_v'\sin\theta_v'\hat{z}],
\end{aligned}
\end{equation}
where $E_{\rm fmw}'$ and $\omega_i'$ are the electric field amplitude and angular frequency of the incident low frequency waves in the comoving frame, respectively.
We perform a Lorentz transformation on the electric field, 
converting it from the comoving frame to the lab frame as
\begin{equation}
\vec E_{\rm rad}=\vec E_{\rm rad,\parallel}'+\gamma(\vec E_{\rm rad,\perp}'-\vec\beta\times\vec B'),
\end{equation}
where the subscripts ``$\parallel$" and ``$\perp$" denote the parallel and perpendicular components of the radiation electric fields with respect to background magnetic field $\vec B_{\rm bg}$.
Then the electric fields of X-mode and O-mode in the lab frame can be generally expressed as
\begin{equation}\label{eq:E_X_ICS}
\begin{aligned}
|\vec E_{\rm X}|&=|\vec E_{\rm rad}\cdot(\hat{n}\times\hat{B}_{\rm bg})|\\
&=|E_x\sin\phi_v\sin\theta_v-E_y\cos\phi_v\sin\theta_v|,
\end{aligned}
\end{equation}
and
\begin{equation}\label{eq:E_O_ICS}
\begin{aligned}
|\vec E_{\rm O}|&=|\vec E_{\rm rad}\times(\hat{n}\times\hat{B}_{\rm bg})|\\
&=|E_z\cos\phi_v\sin\theta_v\hat{i}+E_z\sin\phi_v\sin\theta_v\hat{j}\\
&-(E_x\cos\phi_v\sin\theta_v+E_y\sin\phi_v\sin\theta_v)\hat{k}|,
\end{aligned}
\end{equation}
where the expressions of $E_x$, $E_y$ and $E_z$ are derived in Appendix \ref{app:ICS Lorentz transformation}.
We present the amplitude ratio of X-mode to O-mode waves as a function of the viewing angle ($\theta_v$) in the left panel of Figure~\ref{fig:amplitude ratio} for different azimuthal angles ($\phi_v$). 
One can see that the X- and O-mode amplitudes could be equivalent at specific viewing angles.

Similarly to the case of curvature radiation,
we present the temporal evolution of PA in the right panel of Figure~\ref{fig:amplitude ratio} by considering $\phi_v=60^\circ$.
We note that PA jumps occur within millisecond timescales at the transition of the two orthogonal modes.
As the magnetar spins, the radiation properties will change in the azimuthal direction.
Therefore, we conclude that the ICS mechanism provides a plausible explanation for generating PA jumps via incoherent wave superposition.

\subsubsection{Monster shock}
It has been proposed that fast magnetosonic waves entering the nonlinear region can power different types of high-energy transients, including FRBs, in the closed field line region of magnetars \citep{Beloborodov2023}.
It was suggested that a coherent radio signal may appear ahead of such a monster shock, however, 1D PIC simulations of this physical process have shown that only X-mode waves are generated \citep{Vanthieghem&Levinson2025}.
If this is the case, it is impossible to produce waves with comparable X- and O-modes, so orthogonal PA jumps would rule out such models.

\subsection{A-O-mode conversion}\label{sec:A-O-mode transition}

In the second scenario, we consider an emitter deeper in the magnetosphere where the wave frequency is below the plasma frequency in the comoving frame, i.e. $\omega' < \omega_p'$ (or $r_{\rm em} < r_c$).
The cyclotron frequency of pair plasma is much greater than both $\omega$ and $\omega_p$ inside the magnetar magnetosphere \citep{Arons86}.
In this case, unlike the opposite case where both the O-mode 
and the X-mode waves  
can propagate freely (as discussed in Section \ref{sec:Intrinsic emission mechanisms}), only one of the two orthogonal modes, i.e. the X-mode, or the fast magnetosonic mode (F-mode), can propagate freely. The other mode, i.e. the Alfv\'en mode polarized in the $\vec k-\vec B_{\rm bg}$ plane 
can only propagate along the background magnetic field line.

Suppose that an FRB emitter at an emission radius satisfies $r_{\rm em}<r_c$. Fast magnetosonic waves will propagate freely and convert into X-mode waves with the same emission direction at $r=r_c$, 
Alfv\'en waves, on the other hand, continue change direction until converting themselves into O-mode waves at $r=r_c$, at which they propagate nearly tangentially to the local background magnetic field lines.
Thus, the propagation directions of X-mode and O-mode are different beyond $r>r_c$ (see the lower left panel of Figure~\ref{fig:geometric_cartoon}). 
An observer would see an abrupt change when the emission regions that sweep across the LOS switch from an X-mode-dominated one to an O-mode-dominated one, or vice versa.

When $\omega'/\omega_p$ increases, the plasma oscillation along the background magnetic field is not efficient enough to screen the parallel electric field of the Alfv\'en waves, 
thus A-O-mode conversion occurs and the electric fields of the O-mode waves can have components along the background magnetic field.
The conversion process may happen as the plasma density drops with radius continuously around a characteristic radius.
Around this radius, $\omega'/\omega_p'$ would 
increase slowly from below 
and only approaches unity when the wave becomes a plasma-oscillation Langmuir wave, which would be damped. In order to suddenly cross the $\omega'/\omega_p' = 1$ line and reach the O-mode branch, $\omega'_p$ needs to drop significantly in the distance scale of the wavelength $\lambda'$. We argue that this is marginally possible because the typical size of the plasma bunches that emit FRBs is of the order of $\lambda'$. We envisage that the plasma is composed of these clumps with the characteristic spatial scale of $\lambda'$ in the comoving frame\footnote{Such a configuration is achievable naturally if the FRB emission is powered by inverse Compton scattering off the kilohertz fast magnetosonic waves excited from crustal oscillations \citep{Qu&Zhang2024}.}. At $r_c$, A-O-mode conversion can occur due to the rapid variation of $\omega'/\omega_p'$ around unity within one wavelength scale.

\subsection{Birefringence}\label{sec:birefringence}

When radio waves consisting of two orthogonal modes propagate into a plasma with a distinct refraction index $n > 1$ with an oblique angle, the two modes would propagate in different directions after exiting the plasma, resulting the separation of the two orthogonal modes 
(see the lower right panel of Figure~\ref{fig:geometric_cartoon}).
This effect has been extensively studied within the context of pulsar magnetospheres \citep{Melrose&Stoneham1977,Melrose1979,Allen&Melrose1982,Barnard&Arons1986}.

More generally, for an original FRB wave to undergo refraction, 
the background medium must exhibit spatial variations, and the $n$ gradient must not be parallel to the wave propagation direction (see the lower right panel of Figure~\ref{fig:geometric_cartoon}).

However, the plasma clumps are moving relativistically along $\vec B_{\rm bg}$ and we must consider the $n$ variation in the lab frame. For relativistic motion, 
the refractive index in the lab frame can be calculated as (see Appendix~\ref{app:refractive} for a derivation)
\begin{equation}
n=\frac{\sqrt{\gamma^2\left(n'\cos\theta' + \beta\right)^2 + n'^2\sin^2\theta'}}{\gamma\left(1 + \beta n'\cos\theta'\right)},
\end{equation}
which is $\simeq 1$ for $\gamma\gg1$ and $\theta'\neq1/\gamma$, where $\theta'$ denotes the angle between the wave group velocity and background magnetic field in the comoving frame of the plasma.
In the strong magnetic fields of magnetars, 
the dispersion relations of X-mode and O-mode in the comoving frame of relativistic pair plasma are given by \citep{Arons86}
\begin{equation}
n_{\rm X}'^2=1-\frac{\omega_p'^2}{\omega'^2-\omega_B^2},
\end{equation}
and
\begin{equation}
\begin{aligned}
n_{\rm O}'^2&=\frac{(\omega'^2-\omega_p'^2)(\omega'^2-\omega_p'^2-\omega_B^2)}{(\omega'^2-\omega_p'^2)(\omega'^2-\omega_B^2)-\omega_B^2\omega_p'^2\sin^2\theta_{\rm kB}'}\\
&\simeq\left\{
\begin{aligned}
&n_X'^2, \ &&\theta_{\rm kB}'=0,\\
&1-\frac{\omega_p'^2}{\omega'^2}, \ &&\theta_{\rm kB}'=\pi/2,
\end{aligned}
\right.
\end{aligned}
\end{equation}
where $\theta_{\rm kB}'$ is the angle between wave vector and background magnetic field in the comoving frame.
Note that both X-mode and O-mode waves propagate at speeds close to the speed of light in the open field line region, and their refractive indices are nearly unity.
Thus, plasma birefringence is unlikely to be an important factor to produce PA orthogonal jumps.
We note that plasma birefringence is unlikely to produce PA jumps in the context of FRBs, in contrast to radio pulsars \citep{Barnard&Arons1986}, for the following reasons: FRBs propagate nearly along the background magnetic field in the open field line region where plasma is streaming relativistically outward. This would avoid a significant scattering process from the background leptons \citep{QKZ,Lyutikov2024}.
Additionally, the background magnetic field orientation might be more aligned with FRBs wave vector $\vec k$ since the strong FRB waves are likely to force the plasma to move along $\vec k$ \citep{QKZ}. 
These factors make the birefringence effect less important to produce PA jumps in FRBs.

\subsection{Adiabatic walking and polarization freeze-out}\label{sec:adiabatic-walking}

All the discussion above assumes that the X- and O-mode emissions are those at the emission radius. In a magnetar magnetosphere, a propagation effect known as ``adiabatic walking'' \citep{Cheng&ruderman1979,Lu2019} may operate, which would modify the absolute values of PAs of the two modes. This effect applies to the case that the $\vec k-\vec B_{\rm bg}$ plane and plasma properties slowly vary along the ray path, so that the X- and O-mode waves may undergo PA swing to adjust to the new magnetic configuration. The radius at which such an adiabatic walking stops is called the polarization freeze-out radius or the polarization-limiting radius. \cite{Lu2019} showed that in a magnetar environment and for large-amplitude waves expected in FRBs, the freeze-out radius is within the light cylinder and is approximately the same for both X- and O-modes. This suggests that both wave modes would undergo similar rotations from the emitting radius to the freeze-out radius, so that the relative difference between the PAs of the two modes (which is the one observed) remains the same regardless of the adiabatic walking effect. We therefore do not introduce this effect in the above discussion.

It is worth pointing out that observations of active repeaters with the FAST telescope suggest that the magnetic axis and rotation axis of the magnetar engine are nearly aligned \citep{Luo2025}, see also \cite{Beniamini&Kumar2025}. Under such a condition, 
adiabatic walking is not important for emission from the open field line region far within the light cylinder, because the magnetic field direction essentially does not change.

\section{Conclusions and Discussions}\label{sec:conclusion}

In this paper, we have studied the necessary physical conditions, investigated a variety of intrinsic radiation mechanisms, and propagation effects that may be responsible for the PA jumps observed in three bursts of FRB 20201124A. 
The main conclusions of our study are summarized as follows:
\begin{itemize}
\item In general, PA jumps can arise through the coherent or incoherent superposition of two electromagnetic waves. 
In the coherent superposition case, PA jumps occur when linear polarization reaches a minimum and circular polarization peaks, while conserving the total polarization degree. On the other hand, 
depolarization can occur for incoherent superposition.
Observationally, the three bursts with PA jumps from FRB 20201124A are observed to be depolarized with time, so incoherent superposition is favored.
\item All the processes that could be responsible for PA orthogonal jumps and that have been investigated in this paper are summarized in Figure \ref{fig:map}. 
Our conclusion is that an orthogonal PA jump needs to invoke a geometric effect, 
with different emission regions (within the magnetosphere of the central object) sweeping across the LOS.
Furthermore, the most likely scenario is to invoke intrinsic emission that make both X- and O-mode emissions with comparable amplitudes at different viewing directions. Our critiques for various possibilities are summarized below. 
\item Physically, the millisecond timescales observed from PA jumps impose a stringent constraint on the emission region.
The size of light crossing time $l_{\rm em}=(3\times10^{7} \ {\rm cm})\ \delta t_{-3}$ for a non-relativistic emitter places a tight limit on the emission region size.
The plasma properties and background magnetic field configuration are unlikely to change significantly within millisecond timescales.
Thus, it is difficult for one emitter to produce one orthogonal mode first and then produce another orthogonal mode through a direct physical mechanism outside the magnetosohere. Within the magnetospheric models, rapid PA variations may be generated via physical mechanisms in a dynamically evolving magnetosphere, even though a precise orthogonal mode jump also requires contrived conditions (Figure~\ref{fig:map}). 
\item Geometrically, the upper limit on the probability of PA jumps in FRB 20201124A places a severe constraint on plasma lensing outside the magnetosphere. 
If the observed PA jumps are induced through plasma lensing, the linear number density of the plasma lenses required by observations is $n_l\sim1.5\times10^{-7} \ \rm cm^{-1}$,
which is slightly larger than $n_{\rm F}$, 
and it is unnatural for so many plasma lenses to be located outside the magnetosphere (see the discussion in Section~\ref{sec:general constraint}).
Thus, we rule out plasma lensing outside the magnetosphere as the geometric mechanism producing the PA jumps.
\item Inside the magnetosphere, three intrinsic radiation mechanisms (curvature radiation, inverse Compton scattering, and monster shock model) are investigated. The monster shock model predominantly produces X-mode waves, and orthogonal PA jumps cannot be produced intrinsically.
\item  For curvature radiation from a single emitter
we note that the amplitude ratio of the X-mode to O-mode is viewing angle dependent. 
However, the ratio of $|E_{\rm X}/E_{\rm O}|$ is always greater than unity (see Equation~(\ref{eq:curvature_ratio})). 
Thus, orthogonal PA jumps cannot occur from a single emitter alone and additional propagation effects are required.
\item For the ICS mechanism, 
comparable amplitudes of both X-mode and O-mode waves can be produced at specific viewing angles and azimuthal directions (see Appendix~\ref{app:ICS Lorentz transformation} for a detailed calculation of both modes).
Thus, one extended emitter can produce PA jumps as its different azimuthal directions sweep across the LOS.
The amplitudes of the two orthogonal modes and the temporal evolution of PA are presented in Figure~\ref{fig:amplitude ratio}.
Thus, we conclude that the ICS mechanism can produce orthogonal PA jumps.
\item When FRBs are produced in the deep magnetosphere where the wave frequency is below the plasma frequency in the comoving frame,
as the two MHD waves propagate outward, the background plasma density drops.
At the critical radius $r=r_c$, fast magnetosonic waves converts into X-mode waves. 
A-O-mode conversion occurs if the background plasma density drops in the scale of wavelength, which may be realized for bunched plasmas.
In this case, O-mode waves would propagate nearly in the magnetic field direction at $r=r_c$, which is different from the X-mode direction (see Section~\ref{sec:A-O-mode transition}). 
Thus, the observer would see an abrupt change when an X-mode-dominated zone to an O-mode-dominated zone successively sweeps across the LOS.
\item Inside the magnetosphere, plasma birefringence is less promising, 
as the relativistic motion of the plasma will make the refractive index become nearly unity and plasma birefringence cannot occur.
\end{itemize}

\section*{Acknowledgements}
We thank Kejia Lee, Wenbin Lu and Weiyang Wang for their helpful discussion and an
anonymous referee for helpful comments.
YQ and BZ's work is supported by the Nevada Center for Astrophysics, NASA 80NSSC23M0104 and a Top Tier Doctoral Graduate Research Assistantship (TTDGRA) at University of Nevada, Las Vegas. 
PK’s work was funded in part by an NSF grant AST-2009619 and a NASA grant 80NSSC24K0770.

\appendix

\section{Lorentz transformation of ICS radiation from comoving frame to lab frame}\label{app:ICS Lorentz transformation}
In this Appendix, we present the Lorentz transformation of the ICS radiation from comoving frame to lab frame.
In the comoving frame of the relativistic particle, the radiation electric field $\vec E_{\rm rad}'$ follows Equation~(\ref{eq:single particle's field}). 
One can decompose $\vec E_{\rm rad}'$ into the parallel and perpendicular components along the comoving frame as
\begin{equation}
\vec E_{\rm rad,\parallel}'=\frac{q^2E_{\rm fmw}'\sin(\omega_i't')}{m_ecR'}\left(\frac{\omega_i'}{\omega_B'}\right)\cos\theta_v'\sin\phi_v'\sin\theta_v'\hat{z},
\end{equation}
and
\begin{equation}
\begin{aligned}
\vec E_{\rm rad,\perp}'&=\frac{q^2E_{\rm fmw}'\sin(\omega_i't')}{m_ecR'}\left(\frac{\omega_i'}{\omega_B'}\right)[\cos\phi_v'\sin\phi_v'\sin^2\theta_v'\hat{x}\\
&-(\cos^2\theta_v'+\cos^2\phi_v'\sin^2\theta_v')\hat{y}],
\end{aligned}
\end{equation}
where $\omega_B'=eB_{\rm bg}'/(m_ec)$ is the electron cyclotron frequency in the comoving frame.
We perform Lorentz transformation on the two components to obtain the electric fields in the lab frame as
\begin{equation}\label{eq:E_z}
\vec E_{z}=\vec E_{\rm rad,\parallel}'=\frac{q^2E_{\rm fmw}'\sin(\omega_i't')}{m_ecR'}\left(\frac{\omega_i'}{\omega_B'}\right)\cos\theta_v'\sin\phi_v'\sin\theta_v'\hat{z},
\end{equation}
which is the parallel component along $z$-axis, and 
\begin{equation}
\begin{aligned}
\vec E_{\rm rad,\perp}&=\gamma(\vec E_{\rm rad,\perp}'-\vec\beta\times\vec B')=\gamma[\vec E_{\rm rad,\perp}'-\vec\beta\times(\hat{n}'\times\vec E_{\rm rad}')],
\end{aligned}
\end{equation}
which is the perpendicular component.
We decompose the perpendicular component of electric field $\vec E_{\rm rad,\perp}$ into $x$-axis and $y$-axis components as
\begin{equation}\label{eq:E_x}
E_{x}=\frac{q^2E_{\rm fmw}'\sin(\omega_i't')}{m_ecR'}\left(\frac{\omega_i'}{\omega_B'}\right)\gamma\cos\phi_v'\sin\phi_v'\sin^2\theta_v',
\end{equation}
and
\begin{equation}\label{eq:E_y}
\begin{aligned}
E_{y}&=-\frac{q^2E_{\rm fmw}'\sin(\omega_i't')}{m_ecR'}\left(\frac{\omega_i'}{\omega_B'}\right)\gamma(\cos^2\theta_v'+\beta\cos^3\theta_v'\\
&+\cos^2\phi_v'\sin^2\theta_v'+\beta\cos\theta_v'\sin^2\theta_v'),
\end{aligned}
\end{equation}
which can be submitted into Equations~(\ref{eq:E_X_ICS}) \& (\ref{eq:E_O_ICS}) to calculate the amplitudes of X-mode and O-mode.
The angles can be transformed as $\phi_v'=\phi_v$ and $\sin\theta_v'={\cal D}\sin\theta_v$.

\section{Transformation of refractive index}\label{app:refractive}

In this appendix, we present a brief derivation of the transformation of refractive index in two inertial frames.
The magnitude of the refractive index is defined as $n=|\vec k|c/\omega$.
Thus, the components of the wave vector in the lab frame can be written as
\begin{equation}
k_x = \frac{n\,\omega}{c}\cos\theta, \ k_y = \frac{n\,\omega}{c}\sin\theta.
\end{equation}
The Lorentz transformations of wave vector and angular frequency from the lab frame to the comoving frame which is moving along the $x$-axis are given by
\begin{equation}
k'_x = \gamma\left(k_x - \beta\,\frac{\omega}{c}\right), \quad k'_y = k_y,
\end{equation}
and
\begin{equation}
\omega' = \gamma\left(\omega - \beta c k_x\right).
\end{equation}
The magnitude of the wave vector in the comoving frame of the relativistic medium can be calculated as
\begin{equation}
|\vec{k}'| = \sqrt{(k'_x)^2 + (k'_y)^2} = \frac{\omega}{c}\sqrt{\gamma^2\left(n\cos\theta - \beta\right)^2 + n^2\sin^2\theta}.
\end{equation}
In the comoving frame of the medium, the refractive index $n'$ can be calculated as
\begin{equation}
n'=\frac{c|\vec{k}'|}{\omega'}=\frac{\sqrt{\gamma^2\left(n\cos\theta - \beta\right)^2 + n^2\sin^2\theta}}{\gamma\left(1 - \beta\,n\cos\theta\right)}.
\end{equation}
The inverse transformation rule can be written as
\begin{equation}
n=\frac{\sqrt{\gamma^2\left(n'\cos\theta' + \beta\right)^2 + n'^2\sin^2\theta'}}{\gamma\left(1 + \beta n'\cos\theta'\right)}.
\end{equation}
For a vacuum medium with $n=1$, it follows that $n'=1$ always holds.
When EM waves propagate parallel to the medium moving direction, i.e. $\theta=\theta'=0$, we have
\begin{equation}
n=\frac{n'+\beta}{1+n'\beta}\simeq 1 \ {\rm when} \ \beta\rightarrow1.
\end{equation}
One can see that the relativistic motion of the medium makes the medium more transparent.


\begin{thebibliography}{}
\expandafter\ifx\csname natexlab\endcsname\relax\def\natexlab#1{#1}\fi
\providecommand{\url}[1]{\href{#1}{#1}}
\providecommand{\dodoi}[1]{doi:~\href{http://doi.org/#1}{\nolinkurl{#1}}}
\providecommand{\doeprint}[1]{\href{http://ascl.net/#1}{\nolinkurl{http://ascl.net/#1}}}
\providecommand{\doarXiv}[1]{\href{https://arxiv.org/abs/#1}{\nolinkurl{https://arxiv.org/abs/#1}}}

\bibitem[{{Allen} \& {Melrose}(1982)}]{Allen&Melrose1982}
{Allen}, M.~C., \& {Melrose}, D.~B. 1982, \pasa, 4, 365, \dodoi{10.1017/S1323358000021147}

\bibitem[{{Arons} \& {Barnard}(1986)}]{Arons86}
{Arons}, J., \& {Barnard}, J.~J. 1986, \apj, 302, 120, \dodoi{10.1086/163978}

\bibitem[{{Backer} \& {Rankin}(1980)}]{Backer&Rankin1980}
{Backer}, D.~C., \& {Rankin}, J.~M. 1980, \apjs, 42, 143, \dodoi{10.1086/190647}

\bibitem[{{Backer} {et~al.}(1976){Backer}, {Rankin}, \& {Campbell}}]{Backer1976}
{Backer}, D.~C., {Rankin}, J.~M., \& {Campbell}, D.~B. 1976, \nat, 263, 202, \dodoi{10.1038/263202a0}

\bibitem[{{Barnard} \& {Arons}(1986)}]{Barnard&Arons1986}
{Barnard}, J.~J., \& {Arons}, J. 1986, \apj, 302, 138, \dodoi{10.1086/163979}

\bibitem[{{Beloborodov}(2023)}]{Beloborodov2023}
{Beloborodov}, A.~M. 2023, \apj, 959, 34, \dodoi{10.3847/1538-4357/acf659}

\bibitem[{{Beniamini} \& {Kumar}(2025)}]{Beniamini&Kumar2025}
{Beniamini}, P., \& {Kumar}, P. 2025, \apj, 982, 45, \dodoi{10.3847/1538-4357/adb8e6}

\bibitem[{{Bochenek} {et~al.}(2020){Bochenek}, {Ravi}, {Belov}, {Hallinan}, {Kocz}, {Kulkarni}, \& {McKenna}}]{Bochenek2020}
{Bochenek}, C.~D., {Ravi}, V., {Belov}, K.~V., {et~al.} 2020, \nat, 587, 59, \dodoi{10.1038/s41586-020-2872-x}

\bibitem[{{Chawla} {et~al.}(2020){Chawla}, {Andersen}, {Bhardwaj}, {Fonseca}, {Josephy}, {Kaspi}, {Michilli}, {Pleunis}, {Bandura}, {Bassa}, {Boyle}, {Brar}, {Cassanelli}, {Cubranic}, {Dobbs}, {Dong}, {Gaensler}, {Good}, {Hessels}, {Landecker}, {Leung}, {Li}, {Lin}, {Masui}, {Mckinven}, {Mena-Parra}, {Merryfield}, {Meyers}, {Naidu}, {Ng}, {Patel}, {Rafiei-Ravandi}, {Rahman}, {Sanghavi}, {Scholz}, {Shin}, {Smith}, {Stairs}, {Tendulkar}, \& {Vanderlinde}}]{Chawla2020}
{Chawla}, P., {Andersen}, B.~C., {Bhardwaj}, M., {et~al.} 2020, \apjl, 896, L41, \dodoi{10.3847/2041-8213/ab96bf}

\bibitem[{{Chen} {et~al.}(2022){Chen}, {Yuan}, {Li}, \& {Mahlmann}}]{ChenYR2022}
{Chen}, A.~Y., {Yuan}, Y., {Li}, X., \& {Mahlmann}, J.~F. 2022, arXiv e-prints, arXiv:2210.13506, \dodoi{10.48550/arXiv.2210.13506}

\bibitem[{{Cheng} \& {Ruderman}(1979)}]{Cheng&ruderman1979}
{Cheng}, A.~F., \& {Ruderman}, M.~A. 1979, \apj, 229, 348, \dodoi{10.1086/156959}

\bibitem[{{CHIME/FRB Collaboration} {et~al.}(2019){CHIME/FRB Collaboration}, {Andersen}, {Bandura}, {Bhardwaj}, {Boubel}, {Boyce}, {Boyle}, {Brar}, {Cassanelli}, {Chawla}, {Cubranic}, {Deng}, {Dobbs}, {Fandino}, {Fonseca}, {Gaensler}, {Gilbert}, {Giri}, {Good}, {Halpern}, {Hill}, {Hinshaw}, {H{\"o}fer}, {Josephy}, {Kaspi}, {Kothes}, {Landecker}, {Lang}, {Li}, {Lin}, {Masui}, {Mena-Parra}, {Merryfield}, {Mckinven}, {Michilli}, {Milutinovic}, {Naidu}, {Newburgh}, {Ng}, {Patel}, {Pen}, {Pinsonneault-Marotte}, {Pleunis}, {Rafiei-Ravandi}, {Rahman}, {Ransom}, {Renard}, {Scholz}, {Siegel}, {Singh}, {Smith}, {Stairs}, {Tendulkar}, {Tretyakov}, {Vanderlinde}, {Yadav}, \& {Zwaniga}}]{CHIME2019c}
{CHIME/FRB Collaboration}, {Andersen}, B.~C., {Bandura}, K., {et~al.} 2019, \apjl, 885, L24, \dodoi{10.3847/2041-8213/ab4a80}

\bibitem[{{CHIME/FRB Collaboration} {et~al.}(2020){CHIME/FRB Collaboration}, {Andersen}, {Bandura}, {Bhardwaj}, {Bij}, {Boyce}, {Boyle}, {Brar}, {Cassanelli}, {Chawla}, {Chen}, {Cliche}, {Cook}, {Cubranic}, {Curtin}, {Denman}, {Dobbs}, {Dong}, {Fandino}, {Fonseca}, {Gaensler}, {Giri}, {Good}, {Halpern}, {Hill}, {Hinshaw}, {H{\"o}fer}, {Josephy}, {Kania}, {Kaspi}, {Landecker}, {Leung}, {Li}, {Lin}, {Masui}, {McKinven}, {Mena-Parra}, {Merryfield}, {Meyers}, {Michilli}, {Milutinovic}, {Mirhosseini}, {M{\"u}nchmeyer}, {Naidu}, {Newburgh}, {Ng}, {Patel}, {Pen}, {Pinsonneault-Marotte}, {Pleunis}, {Quine}, {Rafiei-Ravandi}, {Rahman}, {Ransom}, {Renard}, {Sanghavi}, {Scholz}, {Shaw}, {Shin}, {Siegel}, {Singh}, {Smegal}, {Smith}, {Stairs}, {Tan}, {Tendulkar}, {Tretyakov}, {Vanderlinde}, {Wang}, {Wulf}, \& {Zwaniga}}]{CHIME/FRB2020}
{CHIME/FRB Collaboration}, {Andersen}, B.~C., {Bandura}, K.~M., {et~al.} 2020, \nat, 587, 54, \dodoi{10.1038/s41586-020-2863-y}

\bibitem[{{Cocke} \& {Holm}(1972)}]{Cocke&Holm1972}
{Cocke}, W.~J., \& {Holm}, D.~A. 1972, Nature Physical Science, 240, 161, \dodoi{10.1038/physci240161b0}

\bibitem[{{Cordes}(1976)}]{Cordes1976}
{Cordes}, J.~M. 1976, \apj, 210, 780, \dodoi{10.1086/154887}

\bibitem[{{Cordes} \& {Hankins}(1977)}]{Cordes&Hankins1977}
{Cordes}, J.~M., \& {Hankins}, T.~H. 1977, \apj, 218, 484, \dodoi{10.1086/155702}

\bibitem[{{Cordes} {et~al.}(1978){Cordes}, {Rankin}, \& {Backer}}]{Cordes1978}
{Cordes}, J.~M., {Rankin}, J., \& {Backer}, D.~C. 1978, \apj, 223, 961, \dodoi{10.1086/156328}

\bibitem[{{Day} {et~al.}(2020){Day}, {Deller}, {Shannon}, {Qiu(邱昊)}, {Bannister}, {Bhandari}, {Ekers}, {Flynn}, {James}, {Macquart}, {Mahony}, {Phillips}, \& {Xavier Prochaska}}]{Day2020}
{Day}, C.~K., {Deller}, A.~T., {Shannon}, R.~M., {et~al.} 2020, \mnras, 497, 3335, \dodoi{10.1093/mnras/staa2138}

\bibitem[{{Dyks}(2017)}]{Dyks2017}
{Dyks}, J. 2017, \mnras, 472, 4598, \dodoi{10.1093/mnras/stx2101}

\bibitem[{{Dyks}(2019)}]{Dyks2019}
---. 2019, \mnras, 488, 2018, \dodoi{10.1093/mnras/stz1690}

\bibitem[{{Dyks} {et~al.}(2021){Dyks}, {Weltevrede}, \& {Ilie}}]{Dyks2021}
{Dyks}, J., {Weltevrede}, P., \& {Ilie}, C. 2021, \mnras, 501, 2156, \dodoi{10.1093/mnras/staa3762}

\bibitem[{{Er} {et~al.}(2023){Er}, {Pen}, {Sun}, \& {Li}}]{Er2023}
{Er}, X., {Pen}, U.-L., {Sun}, X., \& {Li}, D. 2023, \mnras, 522, 3965, \dodoi{10.1093/mnras/stad1282}

\bibitem[{{Feng} {et~al.}(2022){Feng}, {Li}, {Yang}, {Zhang}, {Zhu}, {Zhang}, {Lu}, {Wang}, {Dai}, {Lynch}, {Yao}, {Jiang}, {Niu}, {Zhou}, {Xu}, {Miao}, {Niu}, {Meng}, {Qian}, {Tsai}, {Wang}, {Xue}, {Yue}, {Yuan}, {Zhang}, \& {Zhang}}]{Feng2022}
{Feng}, Y., {Li}, D., {Yang}, Y.-P., {et~al.} 2022, Science, 375, 1266, \dodoi{10.1126/science.abl7759}

\bibitem[{{Fonseca} {et~al.}(2020){Fonseca}, {Andersen}, {Bhardwaj}, {Chawla}, {Good}, {Josephy}, {Kaspi}, {Masui}, {Mckinven}, {Michilli}, {Pleunis}, {Shin}, {Tendulkar}, {Bandura}, {Boyle}, {Brar}, {Cassanelli}, {Cubranic}, {Dobbs}, {Dong}, {Gaensler}, {Hinshaw}, {Landecker}, {Leung}, {Li}, {Lin}, {Mena-Parra}, {Merryfield}, {Naidu}, {Ng}, {Patel}, {Pen}, {Rafiei-Ravandi}, {Rahman}, {Ransom}, {Scholz}, {Smith}, {Stairs}, {Vanderlinde}, {Yadav}, \& {Zwaniga}}]{Fonseca2020}
{Fonseca}, E., {Andersen}, B.~C., {Bhardwaj}, M., {et~al.} 2020, \apjl, 891, L6, \dodoi{10.3847/2041-8213/ab7208}

\bibitem[{{Gajjar} {et~al.}(2018){Gajjar}, {Siemion}, {Price}, {Law}, {Michilli}, {Hessels}, {Chatterjee}, {Archibald}, {Bower}, {Brinkman}, {Burke-Spolaor}, {Cordes}, {Croft}, {Enriquez}, {Foster}, {Gizani}, {Hellbourg}, {Isaacson}, {Kaspi}, {Lazio}, {Lebofsky}, {Lynch}, {MacMahon}, {McLaughlin}, {Ransom}, {Scholz}, {Seymour}, {Spitler}, {Tendulkar}, {Werthimer}, \& {Zhang}}]{Gajjar2018}
{Gajjar}, V., {Siemion}, A.~P.~V., {Price}, D.~C., {et~al.} 2018, \apj, 863, 2, \dodoi{10.3847/1538-4357/aad005}

\bibitem[{{Hankins} {et~al.}(2003){Hankins}, {Kern}, {Weatherall}, \& {Eilek}}]{Hankins2003}
{Hankins}, T.~H., {Kern}, J.~S., {Weatherall}, J.~C., \& {Eilek}, J.~A. 2003, \nat, 422, 141, \dodoi{10.1038/nature01477}

\bibitem[{{Huang} {et~al.}(2024){Huang}, {Zhong}, \& {Dai}}]{Huang2024}
{Huang}, Y.-C., {Zhong}, S.-Q., \& {Dai}, Z.-G. 2024, \apj, 966, 97, \dodoi{10.3847/1538-4357/ad39e7}

\bibitem[{{Jackson}(1998)}]{Jackson1998}
{Jackson}, J.~D. 1998, {Classical Electrodynamics, 3rd Edition}

\bibitem[{{Jiang} {et~al.}(2022){Jiang}, {Wang}, {Xu}, {Xu}, {Zhang}, {Wang}, {Zhou}, {Zhang}, {Niu}, {Lee}, {Zhang}, {Han}, {Li}, {Zhu}, {Dai}, {Feng}, {Jing}, {Li}, {Luo}, {Miao}, {Niu}, {Tsai}, {Wang}, {Wang}, {Xu}, {Yang}, {Yang}, {Yao}, \& {Yuan}}]{Jiang22}
{Jiang}, J.-C., {Wang}, W.-Y., {Xu}, H., {et~al.} 2022, Research in Astronomy and Astrophysics, 22, 124003, \dodoi{10.1088/1674-4527/ac98f6}

\bibitem[{{Jiang} {et~al.}(2024){Jiang}, {Xu}, {Niu}, {Lee}, {Zhu}, {Zhang}, {Qu}, {Xu}, {Zhou}, {Cao}, {Wang}, {Wang}, {Cao}, {Zhang}, {Zhang}, {Gan}, {Han}, {Hao}, {Huang}, {Jiang}, {Li}, {Li}, {Li}, {Li}, {Luo}, {Men}, {Qian}, {Sun}, {Wang}, {Xu}, {Xu}, {Yang}, {Yao}, {Yue}, {Yu}, {Yuan}, \& {Zhu}}]{Jiang2024}
{Jiang}, J.~C., {Xu}, J.~W., {Niu}, J.~R., {et~al.} 2024, National Science Review, 12, nwae293, \dodoi{10.1093/nsr/nwae293}

\bibitem[{{Kumar} \& {Bo{\v{s}}njak}(2020)}]{Kumar&Bosnjak2020}
{Kumar}, P., \& {Bo{\v{s}}njak}, {\v{Z}}. 2020, \mnras, 494, 2385, \dodoi{10.1093/mnras/staa774}

\bibitem[{{Kumar} {et~al.}(2017){Kumar}, {Lu}, \& {Bhattacharya}}]{Kumar2017}
{Kumar}, P., {Lu}, W., \& {Bhattacharya}, M. 2017, \mnras, 468, 2726, \dodoi{10.1093/mnras/stx665}

\bibitem[{{Kumar} {et~al.}(2021){Kumar}, {Shannon}, {Flynn}, {Os{\l}owski}, {Bhandari}, {Day}, {Deller}, {Farah}, {Kaczmarek}, {Kerr}, {Phillips}, {Price}, {Qiu}, \& {Thyagarajan}}]{PKumar&Shannon2021}
{Kumar}, P., {Shannon}, R.~M., {Flynn}, C., {et~al.} 2021, \mnras, 500, 2525, \dodoi{10.1093/mnras/staa3436}

\bibitem[{{Lorimer} {et~al.}(2007){Lorimer}, {Bailes}, {McLaughlin}, {Narkevic}, \& {Crawford}}]{Lorimer2007}
{Lorimer}, D.~R., {Bailes}, M., {McLaughlin}, M.~A., {Narkevic}, D.~J., \& {Crawford}, F. 2007, Science, 318, 777, \dodoi{10.1126/science.1147532}

\bibitem[{{Lu} {et~al.}(2019){Lu}, {Kumar}, \& {Narayan}}]{Lu2019}
{Lu}, W., {Kumar}, P., \& {Narayan}, R. 2019, \mnras, 483, 359, \dodoi{10.1093/mnras/sty2829}

\bibitem[{{Lu} {et~al.}(2020){Lu}, {Kumar}, \& {Zhang}}]{Lu2020}
{Lu}, W., {Kumar}, P., \& {Zhang}, B. 2020, \mnras, 498, 1397, \dodoi{10.1093/mnras/staa2450}

\bibitem[{{Luo} {et~al.}(2025){Luo}, {Niu}, {Wang}, {Zhang}, {Zhou}, {Xu}, {Wang}, {Niu}, {Zhang}, {Zhang}, {Cai}, {Han}, {Li}, {Lee}, {Zhu}, \& {Zhang}}]{Luo2025}
{Luo}, J.-W., {Niu}, J.-R., {Wang}, W.-Y., {et~al.} 2025, arXiv e-prints, arXiv:2502.16626, \dodoi{10.48550/arXiv.2502.16626}

\bibitem[{{Luo} {et~al.}(2020){Luo}, {Wang}, {Men}, {Zhang}, {Jiang}, {Xu}, {Wang}, {Lee}, {Han}, {Zhang}, {Caballero}, {Chen}, {Chen}, {Gan}, {Guo}, {Hao}, {Huang}, {Jiang}, {Li}, {Li}, {Li}, {Luo}, {Pan}, {Pei}, {Qian}, {Sun}, {Wang}, {Wang}, {Wen}, {Xu}, {Xu}, {Yan}, {Yan}, {Yu}, {Yuan}, {Zhang}, \& {Zhu}}]{Luo2020nature}
{Luo}, R., {Wang}, B.~J., {Men}, Y.~P., {et~al.} 2020, \nat, 586, 693, \dodoi{10.1038/s41586-020-2827-2}

\bibitem[{{Lyutikov}(2024)}]{Lyutikov2024}
{Lyutikov}, M. 2024, \mnras, 529, 2180, \dodoi{10.1093/mnras/stae591}

\bibitem[{{Manchester} {et~al.}(1975){Manchester}, {Taylor}, \& {Huguenin}}]{Manchester1975}
{Manchester}, R.~N., {Taylor}, J.~H., \& {Huguenin}, G.~R. 1975, \apj, 196, 83, \dodoi{10.1086/153395}

\bibitem[{{McKinnon}(2024)}]{McKinnon2024}
{McKinnon}, M.~M. 2024, \apj, 973, 35, \dodoi{10.3847/1538-4357/ad6443}

\bibitem[{{Mckinven} {et~al.}(2024){Mckinven}, {Bhardwaj}, {Eftekhari}, {Kilpatrick}, {Kirichenko}, {Pal}, {Cook}, {Gaensler}, {Giri}, {Kaspi}, {Michilli}, {Nimmo}, {Pearlman}, {Pleunis}, {Sand}, {Stairs}, {Andersen}, {Andrew}, {Bandura}, {Brar}, {Cassanelli}, {Chatterjee}, {Curtin}, {Dong}, {Eadie}, {Fonseca}, {Ibik}, {Kaczmarek}, {Kharel}, {Lazda}, {Leung}, {Li}, {Main}, {Masui}, {Mena-Parra}, {Ng}, {Pandhi}, {Shivraj Patil}, {Prochaska}, {Rafiei-Ravandi}, {Scholz}, {Shah}, {Shin}, \& {Smith}}]{Mckinven2024}
{Mckinven}, R., {Bhardwaj}, M., {Eftekhari}, T., {et~al.} 2024, arXiv e-prints, arXiv:2402.09304, \dodoi{10.48550/arXiv.2402.09304}

\bibitem[{{Melrose}(1979)}]{Melrose1979}
{Melrose}, D.~B. 1979, Australian Journal of Physics, 32, 61, \dodoi{10.1071/PH790061}

\bibitem[{{Melrose} \& {Stoneham}(1977)}]{Melrose&Stoneham1977}
{Melrose}, D.~B., \& {Stoneham}, R.~J. 1977, \pasa, 3, 120, \dodoi{10.1017/S1323358000015010}

\bibitem[{{Metzger} {et~al.}(2019){Metzger}, {Margalit}, \& {Sironi}}]{Metzger2019}
{Metzger}, B.~D., {Margalit}, B., \& {Sironi}, L. 2019, \mnras, 485, 4091, \dodoi{10.1093/mnras/stz700}

\bibitem[{{Michilli} {et~al.}(2018){Michilli}, {Seymour}, {Hessels}, {Spitler}, {Gajjar}, {Archibald}, {Bower}, {Chatterjee}, {Cordes}, {Gourdji}, {Heald}, {Kaspi}, {Law}, {Sobey}, {Adams}, {Bassa}, {Bogdanov}, {Brinkman}, {Demorest}, {Fernandez}, {Hellbourg}, {Lazio}, {Lynch}, {Maddox}, {Marcote}, {McLaughlin}, {Paragi}, {Ransom}, {Scholz}, {Siemion}, {Tendulkar}, {van Rooy}, {Wharton}, \& {Whitlow}}]{Michilli2018}
{Michilli}, D., {Seymour}, A., {Hessels}, J.~W.~T., {et~al.} 2018, \nat, 553, 182, \dodoi{10.1038/nature25149}

\bibitem[{{Nimmo} {et~al.}(2021){Nimmo}, {Hessels}, {Keimpema}, {Archibald}, {Cordes}, {Karuppusamy}, {Kirsten}, {Li}, {Marcote}, \& {Paragi}}]{Nimmo2021}
{Nimmo}, K., {Hessels}, J.~W.~T., {Keimpema}, A., {et~al.} 2021, Nature Astronomy, 5, 594, \dodoi{10.1038/s41550-021-01321-3}

\bibitem[{{Niu} {et~al.}(2024){Niu}, {Wang}, {Jiang}, {Qu}, {Zhou}, {Zhu}, {Lee}, {Han}, {Zhang}, {Li}, {Cao}, {Fang}, {Feng}, {Fu}, {Jiang}, {Jing}, {Li}, {Li}, {Luo}, {Meng}, {Miao}, {Miao}, {Niu}, {Pan}, {Wang}, {Wang}, {Wang}, {Wang}, {Wu}, {Wu}, {Xu}, {Xu}, {Xu}, {Xue}, {Yang}, {Yuan}, {Yue}, {Zhao}, {Zhang}, {Zhang}, {Zhang}, {Zhang}, {Zhang}, \& {Zhu}}]{NiuJR2024}
{Niu}, J.~R., {Wang}, W.~Y., {Jiang}, J.~C., {et~al.} 2024, \apjl, 972, L20, \dodoi{10.3847/2041-8213/ad7023}

\bibitem[{{Oswald} {et~al.}(2023){Oswald}, {Karastergiou}, \& {Johnston}}]{Oswald2023}
{Oswald}, L.~S., {Karastergiou}, A., \& {Johnston}, S. 2023, \mnras, 525, 840, \dodoi{10.1093/mnras/stad2271}

\bibitem[{{Plotnikov} \& {Sironi}(2019)}]{Plotnikov&Sironi2019}
{Plotnikov}, I., \& {Sironi}, L. 2019, \mnras, 485, 3816, \dodoi{10.1093/mnras/stz640}

\bibitem[{{Qu} {et~al.}(2022){Qu}, {Kumar}, \& {Zhang}}]{QKZ}
{Qu}, Y., {Kumar}, P., \& {Zhang}, B. 2022, \mnras, 515, 2020, \dodoi{10.1093/mnras/stac1910}

\bibitem[{{Qu} \& {Zhang}(2024)}]{Qu&Zhang2024}
{Qu}, Y., \& {Zhang}, B. 2024, \apj, 972, 124, \dodoi{10.3847/1538-4357/ad5d5b}

\bibitem[{{Radhakrishnan} \& {Rankin}(1990)}]{Radhakrishnan&Rankin1990}
{Radhakrishnan}, V., \& {Rankin}, J.~M. 1990, \apj, 352, 258, \dodoi{10.1086/168531}

\bibitem[{{Rybicki} \& {Lightman}(1979)}]{Rybicki&Lightman1979}
{Rybicki}, G.~B., \& {Lightman}, A.~P. 1979, {Radiative processes in astrophysics}

\bibitem[{{Singh} {et~al.}(2024){Singh}, {Gupta}, \& {De}}]{Singh2024}
{Singh}, S., {Gupta}, Y., \& {De}, K. 2024, \mnras, 527, 2612, \dodoi{10.1093/mnras/stad3334}

\bibitem[{{Sironi} {et~al.}(2021){Sironi}, {Plotnikov}, {N{\"a}ttil{\"a}}, \& {Beloborodov}}]{Sironi2021}
{Sironi}, L., {Plotnikov}, I., {N{\"a}ttil{\"a}}, J., \& {Beloborodov}, A.~M. 2021, \prl, 127, 035101, \dodoi{10.1103/PhysRevLett.127.035101}

\bibitem[{{Stinebring} {et~al.}(1984){Stinebring}, {Cordes}, {Rankin}, {Weisberg}, \& {Boriakoff}}]{Stinebring1984}
{Stinebring}, D.~R., {Cordes}, J.~M., {Rankin}, J.~M., {Weisberg}, J.~M., \& {Boriakoff}, V. 1984, \apjs, 55, 247, \dodoi{10.1086/190954}

\bibitem[{{Stix}(1992)}]{Stix1992}
{Stix}, T.~H. 1992, {Waves in plasmas}

\bibitem[{{Thornton} {et~al.}(2013){Thornton}, {Stappers}, {Bailes}, {Barsdell}, {Bates}, {Bhat}, {Burgay}, {Burke-Spolaor}, {Champion}, {Coster}, {D'Amico}, {Jameson}, {Johnston}, {Keith}, {Kramer}, {Levin}, {Milia}, {Ng}, {Possenti}, \& {van Straten}}]{Thornton2013}
{Thornton}, D., {Stappers}, B., {Bailes}, M., {et~al.} 2013, Science, 341, 53, \dodoi{10.1126/science.1236789}

\bibitem[{{Vanthieghem} \& {Levinson}(2025)}]{Vanthieghem&Levinson2025}
{Vanthieghem}, A., \& {Levinson}, A. 2025, \prl, 134, 035201, \dodoi{10.1103/PhysRevLett.134.035201}

\bibitem[{{Xu} {et~al.}(2022){Xu}, {Niu}, {Chen}, {Lee}, {Zhu}, {Dong}, {Zhang}, {Jiang}, {Wang}, {Xu}, {Zhang}, {Fu}, {Filippenko}, {Peng}, {Zhou}, {Zhang}, {Wang}, {Feng}, {Li}, {Brink}, {Li}, {Lu}, {Yang}, {Caballero}, {Cai}, {Chen}, {Dai}, {Djorgovski}, {Esamdin}, {Gan}, {Guhathakurta}, {Han}, {Hao}, {Huang}, {Jiang}, {Li}, {Li}, {Li}, {Li}, {Li}, {Liu}, {Luo}, {Men}, {Niu}, {Peng}, {Qian}, {Song}, {Stern}, {Stockton}, {Sun}, {Wang}, {Wang}, {Wang}, {Wang}, {Wu}, {Xiao}, {Xiong}, {Xu}, {Xu}, {Yang}, {Yang}, {Yao}, {Yi}, {Yue}, {Yu}, {Yu}, {Yuan}, {Zhang}, {Zhang}, {Zhang}, {Zhao}, {Zheng}, {Zhu}, \& {Zou}}]{Xu2021}
{Xu}, H., {Niu}, J.~R., {Chen}, P., {et~al.} 2022, \nat, 609, 685, \dodoi{10.1038/s41586-022-05071-8}

\bibitem[{{Zhang}(2022)}]{Zhang22}
{Zhang}, B. 2022, \apj, 925, 53, \dodoi{10.3847/1538-4357/ac3979}

\bibitem[{{Zhang}(2023)}]{Zhang2023RMP}
---. 2023, Reviews of Modern Physics, 95, 035005, \dodoi{10.1103/RevModPhys.95.035005}

\bibitem[{{Zhang} {et~al.}(2023){Zhang}, {Li}, {Zhang}, {Cao}, {Feng}, {Wang}, {Qu}, {Niu}, {Zhu}, {Han}, {Jiang}, {Lee}, {Li}, {Luo}, {Niu}, {Tsai}, {Wang}, {Wang}, {Wu}, {Xu}, {Yang}, {Zhang}, {Zhou}, \& {Zhu}}]{ZhangYK2023}
{Zhang}, Y.-K., {Li}, D., {Zhang}, B., {et~al.} 2023, \apj, 955, 142, \dodoi{10.3847/1538-4357/aced0b}

\end{thebibliography}

\end{document}